\documentclass{article}

\usepackage{arxiv}

\usepackage{amsmath}
\usepackage[utf8]{inputenc} 
\usepackage[T1]{fontenc}    
\usepackage{hyperref}       
\usepackage{url}            
\usepackage{booktabs}       
\usepackage{amsfonts}       
\usepackage{nicefrac}       
\usepackage{microtype}      
\usepackage{cleveref}       
\usepackage{lipsum}         
\usepackage{graphicx}
\usepackage{doi}
\usepackage{booktabs}
\usepackage{comment}
\usepackage[ruled,linesnumbered]{algorithm2e}
\usepackage{subfigure}
\usepackage{caption}
\usepackage{multirow}
\usepackage{listings}
\usepackage{tikz}
\usepackage{enumitem}
\usepackage{tabularx}

\makeatletter
\date{January 2025}           \let\Date\@date
\makeatother

\title{PLANSIEVE: Real-time Suboptimal Query Plan Detection Through Incremental Refinements}

\newif\ifuniqueAffiliation
\uniqueAffiliationtrue

\definecolor{codegreen}{rgb}{0,0.6,0}
\definecolor{codegray}{rgb}{0.5,0.5,0.5}
\definecolor{codepurple}{rgb}{0.58,0,0.82}
\definecolor{backcolour}{rgb}{0.95,0.95,0.92}

\lstdefinestyle{mystyle}{
    backgroundcolor=\color{backcolour},   
    commentstyle=\color{codegreen},
    keywordstyle=\color{purple},
    numberstyle=\tiny\color{codegray},
    stringstyle=\color{codepurple},
    basicstyle=\ttfamily\footnotesize,
    breakatwhitespace=false,         
    breaklines=true,                 
    captionpos=b,                    
    keepspaces=true,                 
    numbers=left,                    
    numbersep=5pt,                  
    showspaces=false,                
    showstringspaces=false,
    showtabs=false,                  
    tabsize=2,
    language=SQL
}

\newcommand*\circled[1]{\tikz[baseline=(char.base)]{
            \node[shape=circle,draw,inner sep=1pt, fill=black, text=white] (char) {#1};}}

\ifuniqueAffiliation 
\author{Asoke Datta, Yesdaulet Izenov, Brian Tsan, Abylay Amanbayev, Florin Rusu\\
	University of California Merced\\
	\texttt{\{adatta2, yizenov, btsan, amanbayev, frusu\}@ucmerced.edu} \\
}
\hypersetup{
pdftitle={PLANSIEVE: Real-time Suboptimal Query Plan Detection Through Incremental Refinements},
pdfauthor={Asoke Datta, Yesdaulet Izenov, Brian Tsan, Abylay Amanbayev, Florin Rusu},
pdfkeywords={query optimization, suboptimal query plans, transformers}
}

\begin{document}
\maketitle

\begin{abstract}
Cardinality estimation remains a fundamental challenge in query optimization, often resulting in sub-optimal execution plans and degraded performance. While errors in cardinality estimation are inevitable, existing methods for identifying sub-optimal plans --- such as metrics like Q-error, P-error, or L1-error --- are limited to post-execution analysis, requiring complete knowledge of true cardinalities and failing to prevent the execution of sub-optimal plans in real-time. This paper introduces PLANSIEVE, a novel framework that identifies sub-optimal plans during query optimization. PLANSIEVE operates by analyzing the relative order of sub-plans generated by the optimizer based on estimated and true cardinalities. It begins with surrogate cardinalities from any third-party estimator and incrementally refines these surrogates as the system processes more queries. Experimental results on the augmented JOB-LIGHT-SCALE and STATS-CEB-SCALE workloads demonstrate that PLANSIEVE achieves an accuracy of up to 88.7\% in predicting sub-optimal plans.
\end{abstract}

\keywords{query optimization, suboptimal query plans, transformers}

\section{INTRODUCTION}\label{sec:intro}
Given a query and a set of enumerated subplans with their respective cardinality estimates, the primary goal of a standard cost-based optimizer is to select an optimal query execution plan. However, cardinality estimates, which are crucial for this process, are prone to error. Inaccurate cardinality estimates can lead to the selection of sub-optimal plans, resulting in significantly longer execution times. As Xiu et al. \cite{Xiu:PARQO:arxiv-2024} demonstrated, even a single subplan of JOB Query Q17 \cite{JOB-github} with an estimation error can increase the actual execution time of the query by up to 5.84x. These estimation errors often stem from the simplifying assumptions that cardinality estimators rely on — such as uniformity of data distribution, independence of predicates, and the principle of inclusion — which frequently fail to represent the intricacies of real-world data \cite{Leis:QOREALLY:pvldb-2015,Leis:JOB:vldb-2018}. Despite decades of research aimed at improving cardinality estimation through better data summaries \cite{Cai:PCETUB:sigmod-2019,Izenov:compass:sigmod-2021, Selinger:APSRDMS:sigmod-1979}, sampling techniques \cite{Leis:index-join-sample:cidr-2017,Muller:ISECKSS:pvldb-2018}, or machine learning models  \cite{Hilprecht:DeepDB:pvldb-2020, Kipf:LCECJDL:cidr-2019, Yang:NeuroCard:pvldb-2021, Kipf:DeepSketches:SIGMOD-2019, Liu:CEUNN:cascon-2015,Malik:BBAQCE:cidr-2007,Ortiz:EADLCE:arxiv-2019,Woltmann:CEL:aiDM-2019,
Reiner:geo-deep:vldb-2024}, the challenge of accurate cardinality estimation remains.

While errors in cardinality estimation are inevitable, efforts can be directed towards mitigating their impact. One significant consequence of such errors is the selection of sub-optimal query execution plans, leading to increased query execution times. Previous research has explored the identification of sub-optimal plans using metrics like the worst-case upper bound \cite{Moerkotte:bad-plans} with P-error \cite{Han:CEB:2021} (the ratio between the selected and optimal plan costs) or L1-error \cite{Izenov:L1-ERROR:sigmod-2024}. However, Izenov et al. \cite{Izenov:L1-ERROR:sigmod-2024} demonstrated that the worst-case upper bound with P-error is a loose metric, lacking a clear correlation with plan sub-optimality. In contrast, they introduced the L1-error metric, which leverages the relative order of subplans based on estimated and true cardinalities. However, both of these metrics are post-analysis tools, requiring access to true cardinalities to identify the optimal plan or calculate the distance between subplan orderings based on estimated and true cardinalities.

Post-execution identification of plan sub-optimality, while valuable for understanding optimizer behavior, does not prevent the execution of a sub-optimal plan in the first place. To proactively avoid the performance penalties associated with such plans, identification needs to occur during the query optimization phase, before execution commences. The challenge lies in the fact that during query optimization, the optimizer lacks access to the true cardinalities of subplans, which are essential for existing methods like the worst-case upper bound with P-error or L1-error. The primary goal of this paper is to address this challenge by developing a framework to identify sub-optimal plans during query optimization time, even without access to true cardinalities.  

In this paper, we introduce PLANSIEVE, a framework designed to identify sub-optimal query plans during the query optimization phase. At the heart of PLANSIEVE is a robust classification model that predicts plan sub-optimality in real-time based on the relative order of subplans. The PLANSIEVE framework operates in two stages: an offline stage and an online stage. In the offline stage, the classification model is trained using subplans true cardinalities to learn the relationship between subplans relative order and plan sub-optimality. In the online phase, PLANSIEVE operates without direct access to true cardinalities, instead utilizing surrogate estimates. As PLANSIEVE processes more queries, it progressively refines these surrogate estimates with the actual cardinalities of subplans observed during query execution.

The primary contributions of this paper includes:
\begin{itemize}[leftmargin=1em, labelindent=1em, itemindent=0em, labelsep=0.5em]
	\item We develop PLANSIEVE, a framework designed to proactively identify sub-optimal query plans during the query optimization phase. As the optimizer enumerates sub-plans and utilizes the system's default cardinality estimator to estimate each subplan's cardinality, PLANSIEVE operates in conjunction with the optimizer to analyze these subplans and their estimates. It constructs two positional vectors: one ordered based on the estimated cardinalities of subplans and another representing the optimal order based on true cardinalities. PLANSIEVE's classification model then determines if the estimated order is likely to lead to a sub-optimal plan. If a potential sub-optimal plan is flagged, the query optimizer can take corrective actions, such as exploring alternative plan options, prioritizing ``safe'' physical operators less sensitive to cardinality estimation errors, or performing targeted subplan estimation correction similar to \cite{Xiu:PARQO:arxiv-2024}. This real-time guidance helps prevent the execution of costly plans, leading to improved query performance and resource utilization. A detailed exposition of the PLANSIEVE framework is provided in Section \ref{sec:framework}.
	
	\item PLANSIEVE utilizes a robust classification model to predict plan sub-optimality. The model employs a transformer architecture to learn positional discrepancies between the subplan order vector based on estimated cardinalities and the optimal order based on true cardinalities. The output of the transformer, which captures these positional discrepancies, is then combined with the calculated L1-error from the subplan order vectors and fed into a Multilayer Perceptron (MLP) for final sub-optimality prediction. This integrated approach, leveraging both positional information and the L1-error metric, enables the classification model to effectively identify sub-optimal query execution plans. An in-depth analysis of the classification model's architecture and its performance evaluation can be found in Sections \ref{sec:classification-model}, \ref{ssec:offline}, and \ref{ssec:online_eval}.

	\item In the online phase, PLANSIEVE initially operates without access to true cardinalities, posing a significant challenge for accurate sub-optimality prediction. To address this, the framework employs surrogate cardinality estimates, serving as temporary placeholders for the actual cardinalities. These surrogates are derived from a third-party cardinality estimator, distinct from the system's default estimator, to ensure their independence from the internally estimated cardinalities. As the host database executes queries, PLANSIEVE diligently collects the true cardinalities of subplans from the execution logs. It employs a multi-faceted cardinality collection strategy, drawing inspiration from prior research \cite{Hertzschuch:physical_operators:vldb_2022, Chaudhuri:true-card}, to cache cardinality data using various subplan patterns. These patterns range from highly selective to less selective, enabling PLANSIEVE to capture a broad spectrum of cardinality distributions. The progressive accumulation of true cardinalities facilitates the refinement of the initial surrogate estimates, leading to continuous improvement in the accuracy of sub-optimality predictions. The specific techniques employed for cardinality collection and refinement are detailed in Section \ref{sec:learn_from_exec}.

	\item The evaluation workloads, JOB-LIGHT and STATS-CEB, while offering a variety of query scenarios, present challenges due to their limited number of queries, which can hinder robust classification model training and evaluation. Furthermore, the JOB-LIGHT workload exhibits a high degree of class imbalance, producing only two sub-optimal plans out of 70 queries. To address these limitations, we have developed a flexible and scalable framework for benchmark augmentation. This augmentation process generates new queries by strategically modifying selection predicates of existing queries, while preserving their core structural characteristics. By leveraging domain knowledge from the original queries used as templates, we ensure that the generated queries remain relevant and realistic. Through this systematic approach, we have significantly expanded the range of query scenarios available by generating 1,357 new queries for JOB-LIGHT and 2,380 for STATS-CEB, resulting in the augmented JOB-LIGHT-SCALE and STATS-CEB-SCALE workloads. Further details on the workload augmentation process are provided in Section \ref{ssec:workload}.
\end{itemize}

\section{PRELIMINARIES}\label{sec:prelim}

This section establishes the key notations used throughout the paper and defines the concept of a ``sub-optimal'' plan within the context of query execution.  We also introduce the L1-error metric, which will serve as a crucial tool in our approach to identifying sub-optimal plans.

\vspace{-0.5em}
\begin{table}[htbp]
    \noindent\rule{\linewidth}{1pt}\vspace{-0.5em}
    \begin{center}
        \textbf{Notations}
    \end{center}
    \vspace{-0.7em}\noindent\rule{\linewidth}{1pt}

    \begin{tabular}{ll} 
    $\boldsymbol{Q}$ & A query \\
    $\boldsymbol{Q_{sub}}$ & Sub-queries of query $\boldsymbol{Q}$ \\
    $\boldsymbol{P}$ & A query plan \\
    $\boldsymbol{P_{opt}}$ & An optimal query plan \\
    $\boldsymbol{P_{sopt}}$ & A sub-optimal query plan \\
    $\boldsymbol{P_{pg}}$ & A query plan generated by DBMS \\
    $\boldsymbol{Q_T}$ & Tables referenced in a query \\
    $\boldsymbol{Q_t}$ & A subset of tables referenced in a query \\
    $\boldsymbol{Y}$ & True cardinality \\
    $\boldsymbol{Y_{s}}$ & True cardinality surrogate \\
    $\boldsymbol{\hat{Y}}$ & The estimated cardinality \\
    $\boldsymbol{N}$ & Number of subplans \\
    $\boldsymbol{k}$ & Join size \\
    $\boldsymbol{N_k}$ & Number of subplans for join size $k$ \\
    $\boldsymbol{\rho}$ & Position vector -- order sorted by true cardinality \\
    $\boldsymbol{\hat{\rho}}$ & Position vector -- order sorted by estimated cardinality \\
    $\boldsymbol{J}$ & Joins in a query \\
    $\boldsymbol{J_k}$ & Joins of size $k$, indicating the number of tables involved \\
    \end{tabular}
    
    \noindent\rule{\linewidth}{1pt}    
\end{table}

\subsection{Problem Statement}
Given a query plan ${P_{pg}}$ produced by an optimizer, which relies on a set of cardinality estimates ${\hat{Y}}$ for its sub-queries ${Q_{sub}}$, our objective is to assess, in real time, whether this plan is sub-optimal. We adopt the definition of plan sub-optimality introduced by Yesdaulet et al. \cite{Izenov:L1-ERROR:sigmod-2024}. A query plan ${P_{pg}}$ is deemed sub-optimal if its cost exceeds the cost an optimal plan ${P_{opt}}$ (calculated with accurate cardinalities ${Y}$) by a factor of (c) or more, where (c) represents a user-specified parameter. The plan derived using true cardinalities ${Y}$ is assumed to be the optimal one. Unlike L1-error, which identify sub-optimal plan ${P_{sopt}}$ in an offline context, PLANSIEVE aims to detect ${P_{sopt}}$ plan during the runtime phase of the query optimization process.

\subsection{L1-Error}
To guide the discussion of various concepts and challenges throughout this paper, we will use the following JOB-LIGHT-SCALE query (\texttt{qry\_68\_9}) as a reference:

\begin{lstlisting}
    SELECT COUNT(*)
    FROM title t, movie_info mi,
    movie_companies mc,
    cast_info ci, movie_keyword mk
    -- selection predicates
    WHERE ci.note in ('(producer)')
    AND mi.info LIKE '%Pacific'
    AND t.production_year > 2000
    -- join predicates
    AND t.id = mi.movie_id
    AND t.id = mc.movie_id
    AND t.id = ci.movie_id
    AND t.id = mk.movie_id
\end{lstlisting}

This query involves five tables : \texttt{movie\_info-mi}, \texttt{title-t},\texttt{movie\_companies-mc}, \texttt{cast\_info-ci}, and \texttt{movie\_keyword-mk}, three selection predicates, and four join predicates. The joins are between primary key/foreign key tables and foreign key/foreign key tables.

\begin{figure*}[htbp]
	\centering
	\includegraphics[width=\textwidth]{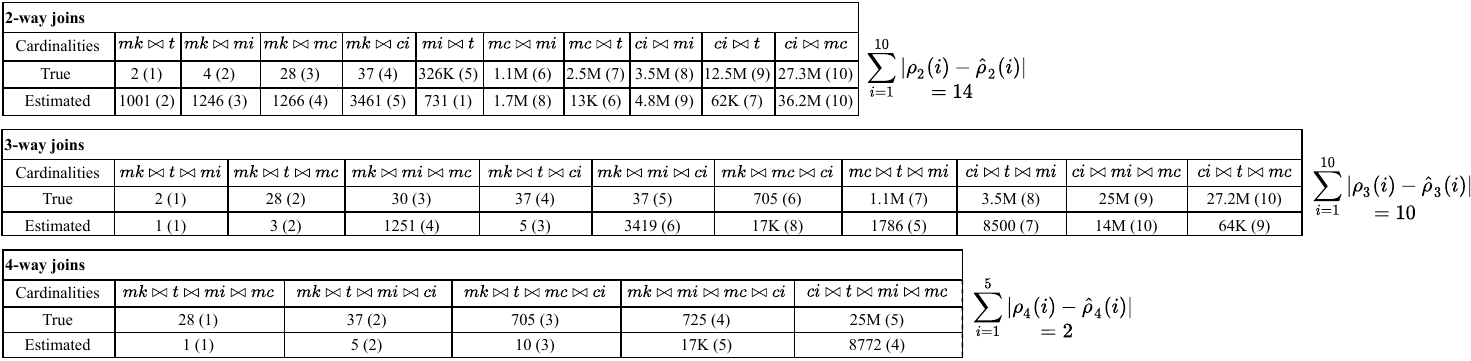}
	\caption{L1-error calculation for query qry\_68\_9.}
	\label{fig:all_subs}
\end{figure*}

During the exploration of the plan space for query \texttt{qry\_68\_9}, the optimizer enumerates a set of subqueries, denoted as  \texttt{$Q_{sub}$} (Figure \ref{fig:all_subs}), which includes various structures such as left-deep, bushy, or others as defined by the database optimizer, which may differ across databases. The objective is to determine if the generated plan, based on these subplans and their estimated cardinalities, will be sub-optimal. To quantify plan sub-optimality, Izenov et al. \cite{Izenov:L1-ERROR:sigmod-2024} propose the L1-error metric, which assesses the positional discrepancies between the subplan orderings derived from true cardinalities (${\rho}$) and estimated cardinalities (${\hat{\rho}}$). The premise is that larger positional differences correlate with a higher likelihood of sub-optimal plan generation, while the absence of such differences suggests an optimal plan. The L1-error can be computed using either the costs of physical operations or solely based on cardinality. PLANSIEVE leverages this insight, incorporating the L1-error as a pivotal feature in its machine learning model. 

The computation of the L1-error for a query involves a two-step process. First, the L1-error is calculated for each join size $k$. These individual L1-errors are then aggregated to obtain the overall L1-error for the query. For a specific join size $k$, the process entails sorting the subplans  $Q_{sub}$ in ascending order of their true cardinalities $Y$, yielding the position vector $\rho$. The subplans are then sorted based on their estimated cardinalities $\hat{Y}$, resulting in the position vector $\hat{\rho}$. The L1-error for join size $k$ is then computed by summing the absolute differences between the corresponding positions in $\rho$ and $\hat{\rho}$ is given by:

\begin{equation}
    \sum_{i=1}^{{N}_{k}}|\rho(i) - \hat{\rho}(i)|
\end{equation}

where ${N}_{k}$ represents the number of subplans with join size $k$. Illustratively, in Figure \ref{fig:all_subs}, for 2-way joins (k = 2), the subplan ($mk \Join t$) ranks 1st in the true cardinality-based order but 2nd in the estimated order, resulting in a positional difference of $|1 - 2| = 1$. Additionally, reversing the join order in subplans, such as changing ($mi \Join t$) to ($t \Join mi$), would yield the same cardinality. However, if physical operators are considered, the resulting value may vary. For the subplan ($mi \Join t$), the positional difference is $|5 - 1| = 4$. The summation of these differences across all subplans of a particular join size yields the total positional difference for that join size. In the case of 2-way joins (k = 2), this sum amounts to 14. For 3-way and 4-way joins, the total positional differences are computed to be 10 and 2, respectively.

To arrive at the query-level L1-error, the individual join size L1-errors are combined, with a weighting scheme that prioritizes lower-level joins. This weighting scheme reflects the cascading impact of cardinality estimation errors in early join stages on the overall query execution cost. Consequently, errors in 2-way joins contribute more significantly to the total L1-error than those in 4-way joins, ensuring that the metric accurately captures the potential ramifications of estimation inaccuracies at different stages of the query plan.

\section{PLANSIEVE FRAMEWORK}\label{sec:framework}
PLANSIEVE tackles the persistent challenge of identifying sub-optimal query plans, ${P_{sopt}}$, in real-time. It achieves this by combining a sophisticated machine learning model with a mechanism to dynamically collect and leverage true cardinalities, ${Y}$, from historical query executions. This enables the proactive detection and potential correction of inefficient plans during the query optimization process itself, before execution commences. The subsequent subsections will elaborate on the framework's architecture, workflow, and potential applications.

\begin{figure*}[htbp]
	\centering
	\includegraphics[width = .85\textwidth]{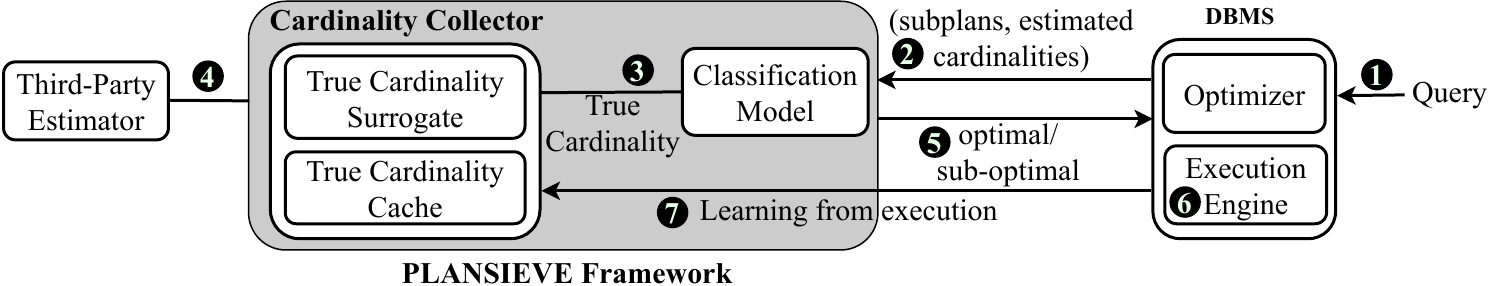}
	\caption{PLANSEIVE framework architecture.}
	\label{fig:plansieve}
\end{figure*}

\subsection{Architecture}
The PLANSIEVE framework, as illustrated in Figure \ref{fig:plansieve}, centers around three primary components that collaborate to identify sub-optimal query plans: DBMS, the Classification Model and the Cardinality Collector.

\subsubsection{DBMS}
PLANSIEVE integrates with a DBMS to collect and analyze query execution plans, and while it can be adapted to various systems, our implementation focuses on the open-source PostgreSQL database \cite{postgres} due to its accessible optimizer. Upon receiving a query ${Q}$, the optimizer initiates its plan enumeration process, exploring the possible execution strategies within the search space. During this process, PLANSIEVE extracts the enumerated subqueries ${Q_{\text{sub}}}$ and their corresponding cardinality estimates ${\hat{Y}}$ directly from the optimizer.

The number of subqueries generated depends on both the complexity of the query ${Q}$ and the breadth of the optimizer's search space. For instance, consider the query \texttt{qry\_68\_9}, where the optimizer's enumerated subqueries are shown in Figure \ref{fig:all_subs},  grouped by their join size $k$. This query, involving four joins across five tables, results in subqueries with join sizes ranging from two to four. Notably, the optimizer also evaluates 
subqueries like $mk \Join mi$ or $mk \Join mc$, even though these joins are not explicitly stated in the original query. This behavior arises from the optimizer's use of transitivity properties in join predicates. For example, with joins between $t$ and $mi$ (t.id = mi.movie\_id) and between t and mk (t.id = mk.movie\_id), the optimizer infers and evaluates the potential join mi.movie\_id = mk.movie\_id. This sophisticated handling of transitive joins highlights the optimizer's thorough exploration of the search space to identify the most optimal execution plans.

\subsubsection{Classification Model}
The Classification Model serves as the predictive core of PLANSIEVE. This sophisticated machine learning model, detailed in section \ref{sec:classification-model}, is designed to assess the optimality of a query plan ${P}$, particularly focusing on the potential sub-optimality arising from inaccuracies in the estimated cardinalities, ${\hat{Y}}$, of its constituent subplans, ${Q_{sub}}$. The model tackles the inherent challenges posed by the varying number of subplans, ${N}$, across queries and the critical importance of their relative order within the plan. It achieves this by employing a transformer-based architecture, adapted from the GPT-2 model commonly used in natural language processing tasks. This adaptation empowers the model to effectively process the variable-length sequences of subplans, accommodating the diverse range of query scenarios encountered in real-world databases.

The transformer's self-attention mechanism plays a pivotal role in capturing the intricate relationships and dependencies between subplans within the sequence. This enables the model to learn the subtle nuances of how the ordering of subplans, represented by the position vectors ${\rho}$ (true cardinality) and ${\hat{\rho}}$ (estimated cardinality), influences the overall plan optimality. The transformer's output, enriched with the aggregated L1-error --- a measure of the discrepancy between ${\rho}$ and  ${\hat{\rho}}$ -- is then fed into a Multilayer Perceptron (MLP) for classification. The MLP, equipped with non-linear activation functions, learns complex decision boundaries, enabling it to accurately predict whether a plan, P, is likely to be optimal (${P_{opt}}$) or sub-optimal (${P_{sopt}}$).

\subsubsection{Cardinality Collector}
The Cardinality Collector, detailed in Section \ref{sec:learn_from_exec}, bridges the gap between estimated cardinalities, ${\hat{Y}}$, and true cardinalities, ${Y}$ , of subplans ${Q_{sub}}$, playing a crucial role in enabling PLANSIEVE's real-time predictions. It maintains a cache of true cardinalities observed from past query executions, facilitating efficient retrieval when encountering familiar subplans.

In the case of a cache miss (i.e., a new or unseen subplan), the Cardinality Collector employs a third-party cardinality estimator to provide a surrogate estimate, ${Y_{s}}$. This third-party estimator can be a single model or an ensemble of multiple estimators, carefully selected to align with the specific data and workload characteristics. As the system processes more queries, the cache is dynamically enriched with the actual cardinalities, ${Y}$, obtained from query execution. This ongoing refinement of initial ${Y_{s}}$ leads to progressively more accurate predictions, ensuring PLANSIEVE's adaptability to the evolving data landscape and shifting query patterns.

\subsubsection{Third-Party Estimator}
PLANSIEVE incorporates a third-party cardinality estimator to generate surrogate cardinalities, ${Y_{s}}$, offering flexibility in estimator selection. It can integrate any suitable estimator or an ensemble of methods, accommodating both traditional and machine learning-based models, to mitigate bias and adapt to diverse data environments. This design choice enhances PLANSIEVE's adaptability to various query workloads and data characteristics.

\subsection{Workflow}
The workflow of PLANSIEVE is meticulously designed to identify sub-optimal query plans in real-time, leveraging a set of integrated components that function in unison. \circled{1} Upon receiving a query, ${Q}$, the Query Optimizer begins by enumerating potential subplans, ${Q_{sub}}$, and estimating their cardinalities, ${\hat{Y}}$. \circled{2} These ${Q_{sub}}$ coupled with their ${\hat{Y}}$ , are then passed to the Classification Model, which plays a pivotal role in assessing whether a given plan is likely to be sub-optimal.

The Classification Model's ability to make accurate predictions depends on its interaction with the Cardinality Collector, which provides the true cardinality, ${Y}$, values \circled{3} necessary for precise evaluation. The Cardinality Collector first attempts to retrieve ${Y}$ values from the True Cardinality Cache. However, in the event of a cache miss, it utilizes a third-party estimator to generate surrogate, ${Y_{s}}$, values \circled{4}. This ensures that the Classification Model has the comprehensive data to construct the true cardinality-based position vector ${\rho}$ and accurately calculate the L1-error. These elements are critical for assessing the potential sub-optimality of a query plan.

Based on the Classification Model's prediction, the Query Optimizer receives feedback that informs its subsequent decisions \circled{5}. If a plan is flagged as sub-optimal, the optimizer may either select an alternative plan or modify its strategy in response to the insights provided by PLANSIEVE. If a query proceeds to execution, the Execution Engine executes the plan \circled{6}, and the resulting execution log --- such as the result of the \texttt{EXPLAIN ANALYZE} command in PostgreSQL, which includes actual cardinality data --- is fed back into the Cardinality Collector \circled{7}. This log is instrumental in updating the cache, allowing PLANSIEVE to refine its predictions continuously. As PLANSIEVE processes more queries, this feedback loop enhances its accuracy, enabling the system to more effectively predict and avoid sub-optimal plans in future queries.

\subsection{Implementation}
The PLANSEIVE framework, as illustrated in Figure \ref{fig:plansieve}, seamlessly integrates with a database management system (DBMS) to facilitate the real-time identification of suboptimal query plans ${P_{sopt}}$. Our implementation leverages the open-source PostgreSQL database, capitalizing on its accessible optimizer to extract enumerated subplans ${Q_{sub}}$ and their estimated cardinalities ${\hat{Y}}$ during query optimization. In scenarios where true cardinalities $Y$ are unavailable in cache, PLANSEIVE employs DeepDB \cite{Hilprecht:DeepDB:pvldb-2020}, a third-party cardinality estimator utilizing machine learning techniques, to generate surrogate estimates ${Y_{s}}$. The framework also employs the L1-error metric \cite{Izenov:L1-ERROR:sigmod-2024} to quantify discrepancies between estimated and actual cardinalities, serving as a robust indicator for assessing potential plan sub-optimality.

The core of PLANSEIVE lies in its meticulously designed classification model, implemented using PyTorch. The model leverages a transformer architecture, specifically GPT2 Sequence Classification, to capture the nuanced relationships between subplan positions within a query plan $P$. Following the transformer, a multilayer perceptron (MLP) performs the final classification, determining the likelihood of a query plan P being suboptimal ${P_{sopt}}$ based on subplan ordering and cardinality discrepancies.

Furthermore, PLANSEIVE incorporates a sophisticated caching mechanism, utilizing a key-value map data structure, to efficiently store and retrieve subplans ${Q_{sub}}$ and their associated true cardinalities $Y$. This mechanism not only enables rapid access to previously computed values but also supports the Cardinality Collector's continuous learning from historical query executions, refining its understanding of query patterns and subplan cardinalities over time.

\subsection{Applications}
PLANSEIVE's ability to identify sub-optimal query plans, ${P_{sopt}}$, in real-time opens up several potential applications for improving query optimization and overall database performance:

\noindent
\textbf{Guiding Query Optimizer Decisions.} When PLANSIEVE flags a plan $P$ as potentially sub-optimal, it provides the query optimizer with valuable information to guide subsequent actions. The optimizer could explore alternative plan options, prioritize the use of ``safe'' physical operators known to be less sensitive to cardinality estimation errors, or even trigger query re-optimization based on the model's feedback. This real-time guidance can help avoid the execution of costly plans, leading to improved query performance and resource utilization.

\noindent
\textbf{Targeted Estimate Correction.} PLANSIEVE's focus on the relative order of enumerated subplans, ${Q_{sub}}$, offers a unique opportunity for targeted estimate correction. By comparing the estimated cardinality-based position vector ${\hat{\rho}}$ and the true cardinality-based position vector ${\rho}$, PLANSIEVE can pinpoint specific subplans whose estimates ${\hat{Y}}$ are significantly out of sync with reality ${Y}$. These problematic estimates can then be flagged for re-estimation or refinement, potentially leading to improved plan choices in subsequent queries.

These diverse applications highlight the versatility and potential impact of PLANSIEVE in enhancing query optimization. Its real-time, data-driven approach can not only improve immediate query performance but also contribute to long-term optimization strategies, making it a valuable tool for database administrators and researchers alike.

\section{CARDINALITY COLLECTOR}\label{sec:learn_from_exec}

This section provides a comprehensive overview of the Cardinality Collector module within PLANSIEVE, which is tasked with acquiring true cardinalities ${Y}$ from query execution and refining initial surrogate estimates ${Y_{s}}$ to accurately assess the sub-optimality of query plans. We will explore how PLANSIEVE manages situations where true cardinalities ${Y}$ are initially unavailable, addressing cache misses by leveraging surrogate estimates ${Y_{s}}$ from third-party estimators. Additionally, the section delves into the selection and integration of these estimators, whether used individually or in an ensemble, to enhance accuracy and minimize bias. The process of collecting true cardinalities ${Y}$ via execution logs, strategies for caching these values, and approaches for handling partial matches for sub-plans ${Q_{sub}}$ are also discussed. This framework ensures that PLANSIEVE continually improves its predictions as more queries are processed, thereby enhancing its ability to identify suboptimal query plans ${P_{sopt}}$ in real-time.

\begin{figure}[htbp]
	\centering
	\includegraphics[width = .3\linewidth]{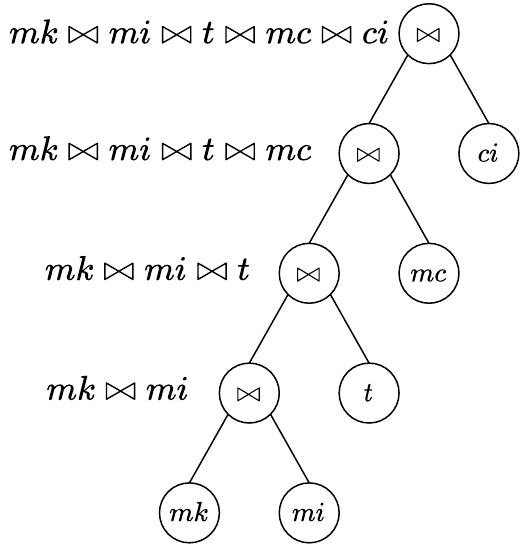}
	\caption{Execution plan for query \texttt{t\_68\_9}.}
	\label{fig:exec_plan}
\end{figure}

\subsection{Cache Misses}
To construct a comprehensive feature representation for queries, as outlined in Section \ref{sec:classification-model}, and to accurately calculate the L1-error, the classification model requires complete knowledge of both the true cardinalities ${Y}$ and the estimated cardinalities ${\hat{Y}}$ for all considered sub-plans ${Q_{sub}}$ within a query plan ${P}$. However, obtaining the exact cardinalities ${Y}$ necessitates executing all sub-plans ${Q_{sub}}$ and caching their results, which is not feasible in an online setting.

When PLANSIEVE processes a new query ${Q}$ from a workload and the true cardinalities ${Y}$ for its sub-plans ${Q_{sub}}$ are not available, we encounter a scenario known as a cache miss. To address this challenge, PLANSIEVE initially relies on a third-party cardinality estimator to generate surrogate estimates, ${Y_{s}}$, for these sub-plans. These surrogate estimates ${Y_{s}}$ serve as a temporary stand-in for the true cardinalities. As PLANSIEVE continues to process more queries and their actual cardinalities ${Y}$ are gathered, these initial surrogate estimates ${Y_{s}}$ are progressively refined, thereby enhancing both the granularity and accuracy of the system over time.

PLANSIEVE's design allows for the deployment of either a single cardinality estimator or an ensemble of multiple estimators. The use of an ensemble is particularly advantageous as it helps to mitigate potential biases that might arise from overestimation or underestimation by leveraging the diverse strengths of multiple models. This flexibility ensures that PLANSIEVE can adapt to varying data characteristics and workload patterns, providing a robust and reliable estimation process for surrogate estimates, ${Y_{s}}$.

In our implementation with PostgreSQL, we have deliberately employed a third-party estimator distinct from PostgreSQL's native cardinality estimation mechanism. This strategic choice ensures that the true cardinality surrogates ${Y_{s}}$ remain independent of the internally estimated cardinalities ${\hat{Y}}$, preventing potential bias. By maintaining this separation, PLANSIEVE accurately differentiates between the DBMS's native estimates ${\hat{Y}}$ and the true cardinality surrogates ${Y_{s}}$, which is crucial for precise L1-error calculation and reliable assessment of query plan sub-optimality initially.

\subsubsection{Third-Party Estimator} 
To effectively manage the diversity of query workloads and the varying data characteristics encountered in real-world scenarios, PLANSIEVE integrates third-party cardinality estimators with a judicious and strategic approach. Rather than relying on a one-size-fits-all solution, PLANSIEVE selectively employs specific models --- or an ensemble of models --- that best align with the performance demands and data environment at hand. This careful selection process is informed by a thorough evaluation of each estimator’s strengths and their suitability for the task, ensuring that the chosen estimators complement one another and collectively enhance the accuracy of surrogate cardinality estimates, $\boldsymbol{Y_{s}}$.

Third-party cardinality estimation methods span a broad spectrum, from traditional synopses-based techniques to advanced machine learning models. Traditional methods, such as statistical synopses, leverage histograms \cite{postgres}, sampling \cite{Leis:index-join-sample:cidr-2017,Muller:ISECKSS:pvldb-2018}, and sketches \cite{Izenov:compass:sigmod-2021,Cai:PCETUB:sigmod-2019,Rusu:SAS:sigmod-2007, Rusu:SJSE:tods-2008} to efficiently estimate data distributions. These approaches, while robust in many cases, may struggle to capture the complex patterns and correlations inherent in modern, diverse datasets --- limitations that machine learning techniques are often better equipped to address.

Machine learning-based estimators can be categorized into two primary groups: those trained directly on data distributions and those tailored to specific workload patterns. The former, as exemplified by models like DeepDB \cite{Hilprecht:DeepDB:pvldb-2020, Wu:BayesCard:arXiv, Yang:naru:VLDB-2019, Yang:NeuroCard:pvldb-2021, Woltmann:CEL:aiDM-2019} focus on learning the underlying characteristics of the data itself, providing a generalized approach capable of adapting to a wide range of queries. In contrast, workload-trained models, such as those discussed in \cite{Kipf:LCECJDL:cidr-2019, Negi:RQC:PVLDB-2023, Marcus:Bao:SIGMOD-2021}, derive their predictions from historical query patterns, offering high accuracy for queries similar to those seen in the training phase but potentially faltering when faced with novel or altered query structures.

By harnessing the strengths of these varied methodologies, PLANSIEVE enhances its ability to generate reliable surrogate cardinalities ${Y_{s}}$, ultimately improving its capacity to detect suboptimal query plans ${P_{sopt}}$ in real-time.

\begin{table*}[htbp]
    \centering
    \begin{tabular}{clllr}
    \toprule
     & \textbf{Selection-Sensitive Pattern} & \textbf{Pattern (Approach 1)} & \textbf{Pattern (Approach 2)} & \textbf{Cardinality} \\ \midrule
    Q1 & \(mk \Join mi_{\sigma_{\text{info LIKE '\%Pacific'}}} \Join t_{\sigma_{\text{production\_year} > 2000}}\) & \(mk \Join mi \Join t\) & \(mk \Join mi_{\sigma} \Join t_{\sigma}\) & 25.7K \\ 
    Q2 & \(mk \Join mi_{\sigma_{\text{info LIKE '\%Pacific'}}} \Join t_{\sigma_{\text{production\_year} > 1990}}\) & \(mk \Join mi \Join t\) & \(mk \Join mi_{\sigma} \Join t_{\sigma}\) & 35.3K\\ 
    Q3 & \(mk \Join mi_{\sigma_{\text{info LIKE '\%Pacific'}}} \Join t\) & \(mk \Join mi \Join t\) & \(mk \Join mi_{\sigma} \Join t\) & 58.3K \\ 
    Q4 & \(mk \Join mi \Join t_{\sigma_{\text{production\_year} > 1990}}\) & \(mk \Join mi \Join t\) & \(mk \Join mi \Join t_{\sigma}\) & 157M\\
    Q5 & \(mk \Join mi \Join t\) & \(mk \Join mi \Join t\) & \(mk \Join mi \Join t\) &  235.4M\\ \bottomrule 
    \end{tabular}
    \caption{Cardinality estimation pattern types.}
    \label{tab:query_patterns}
\end{table*}

\subsection{Collecting True Cardinalities}
When executing a query ${Q}$, an execution log is generated --- using PostgreSQL's \texttt{EXPLAIN ANALYZE} command --- that provides detailed insights into the actual cardinality ${Y}$ of the entire query, as well as the cardinalities of its constituent sub-plans ${Q_{sub}}$, including the selection predicate cardinalities for individual tables, as highlighted by Chaudhuri et al. \cite{Chaudhuri:true-card}. These cardinality data ${Y}$ are crucial for subsequent decision-making processes. The cardinality values ${Y}$ for individual tables can be utilized to construct histograms using workload-aware histogram building techniques, as discussed in the literature \cite{ashraf:self-tuning-hist:sigmod-1999,bruno:stholes:sigmod-2021}. While these histograms are effective for approximating single table cardinality in future queries, they do not scale well for scenarios involving joins.

Given that PLANSIEVE focuses on the cardinality of join sub-plans, the Cardinality Collector in PLANSIEVE systematically records these sub-plans and their corresponding cardinalities as key-value pairs within its cache. The subplans for \texttt{qry\_68\_9} are depicted in Figure \ref{fig:exec_plan}.  The cardinality collector records these subplans and their cardinality as a pair --- [subplan pattern, cardinality]. For example, subplan $mk \Join mi$ is reprsented as [$mk\Join {mi}_{\sigma_{\text{info LIKE '\%Pacific'}}}, 4$], subplan $mk \Join mi \Join t$ is reprsented as [$mk\Join {mi}_{\sigma_{\text{info LIKE '\%Pacific'}}} \Join {t}_{\sigma_{\text{production\_year > 2000}}}, 2$] in cache. These recorded pairs allow PLANSIEVE to refine its true cardinality surrogates ${Y_{s}}$, improving its overall effectiveness in detecting sub-optimal query plans.

\noindent
\textbf{Cache Hit and Pattern Building for Partial Matching.}\label{ssec:cache-hit}
When an exact pattern match for a sub-plan ${Q_{sub}}$ is found in the cache, the Cardinality Collector immediately retrieves it, a scenario referred to as a Cache Hit. However, when an exact match is unavailable, the Cardinality Collector searches for a partial match. A partial match occurs when the join conditions between sub-plans ${Q_{sub}}$ are identical, but the selection predicates differ. For instance, Table  \ref{tab:query_patterns} illustrates five different patterns that are considered partial matches of each other, where variations in selection predicates may impact one or more tables.

To effectively handle partial matches, the Cardinality Collector employs various strategies for pattern creation. The following paragraphs outline the methods considered by the Cardinality Collector when constructing patterns for partial matching. Specifically, we discuss two different approaches to constructing these patterns.

\underline{\textbf{Approach 1.}} Patterns derived during true cardinality $Y$ parsing are sensitive to selections, known as selection-sensitive patterns. These patterns do not scale well in workloads involving numerous selection predicates. To address this limitation, Approach 1 proposes the creation of more generic patterns. As illustrated in Table \ref{tab:query_patterns}, since all subplans share the same join predicates, they form the same generic pattern --- $mk \Join mi \Join t$. 
\[
\sum_{1}^{N}{Y}_{\textit{p}} = \frac{{Y}_{{\textit{p}_{1}}} + {Y}_{{\textit{p}_{2}}} + {Y}_{{\textit{p}_{3}}} + .... + {Y}_{{\textit{p}_{n}}}}{n}
\]
If a pattern exists in the cache, cardinalities are aggregated in a method akin to that used in TONIC \cite{Hertzschuch:physical_operators:vldb_2022} using the above formula --- ${Y}_{{\textit{p}_{n}}}$ is the most recent sub plan's cardinality --- and averaged. This averaged cardinality value is subsequently used to refresh the cache, effectively replacing the previous generic pattern's value. Additional emphasis can be given to the most current cardinality values to better reflect recent changes in the workload or data.

\underline{\textbf{Approach 2.}} Approach 1 treats all selection-sensitive patterns equally, as depicted in Table \ref{tab:query_patterns}, because they share the same join conditions. While it introduces less memory overhead, it may not accurately reflect the wide variance in Cardinality --- ranging from 25.7K to 235.4M in Table \ref{tab:query_patterns}. An average cannot effectively represent the distribution in such a broad spectrum of values. Approach 2 addresses this issue by creating a selection-aware pattern based on the presence of selection predicates. It does not record the exact predicate details but notes the presence of selection conditions. For example, in this approach, Q1 and Q2 yield the same pattern --- $mk \Join mi_{\sigma} \Join t_{\sigma}$ --- due to having selection conditions on identical tables. Conversely, Q3, Q4, and Q5 each create distinct patterns in Approach 2, reflecting their unique selection contexts. Cardinalities for equivalent patterns can be combined as in Approach 1, or alternatively, the distribution of cardinalities for patterns can be explored over time. To support this gradual analysis, methods such as reservoir sampling or estimators like kernel density estimators (KDE) and Gaussian Mixture Models (GMM) can be utilized to incrementally assess the cardinality distribution.

\section{CLASSIFICATION MODEL}\label{sec:classification-model}

Given a query ${Q}$, its enumerated subplans ${Q_{sub}}$, along with their true cardinality-based position vectors ${\rho}$ and estimated cardinality-based position vectors ${\hat{\rho}}$, and the corresponding query-level L1-error, the classification model is designed to learn the positional discrepancies between these vectors within a supervised learning framework. The model's input consists of the tuple (${\rho}$, ${\hat{\rho}}$, L1-error), and the output is a binary label indicating whether the query plan ${P}$, generated using the subplans estimated cardinalities ${\hat{Y}}$, is suboptimal or optimal. This label is determined by evaluating the P-error, which quantifies the cost ratio between the plan based on estimated cardinalities ${\hat{Y}}$ and the optimal plan based on true cardinalities ${Y}$.

The primary challenges in constructing an effective classification model lie in the representation of queries and the selection of a suitable model architecture. The core input to our model comprises two positional vectors, ${\rho}$ and ${\hat{\rho}}$, representing distinct orderings of the same set of subplans, ${Q_{sub}}$. Ensuring that the model can effectively differentiate and process these two sequences independently poses a significant hurdle. Furthermore, the number of subplans, $N$, is inherently variable, fluctuating within a single query based on the join size $k$, and differing across queries depending on the tables ${Q_T}$ or joins ${J}$ involved. This variability necessitates a meaningful representation of query subplans and the selection of a model architecture capable of accommodating inputs of varying lengths.

A further challenge lies in the acquisition of the initial training dataset, often hindered by the ``cold start'' problem - the lack of sufficient historical query execution data to train the model effectively. In the subsequent discussion, we address these foundational challenges by elucidating strategies for effective query representation and initial data collection. We will then provide a concise overview of the model architecture, emphasizing its capability to integrate these inputs to accurately assess the sub-optimality of query plans $P$.

\subsection{Query Representation}\label{ssec:query_representation}

Each query ${Q}$ is represented through its subplans ${Q_{sub}}$, encoded using two positional vectors for each join size ${k}$: ${\rho_k}$ and ${\hat{\rho}_k}$. The vector ${\rho_k}$ captures the relative ordering of subplans when sorted by their true cardinalities ${Y}$, while ${\hat{\rho}_k}$ represents the order based on estimated cardinalities ${\hat{Y}}$. According to Izenov et al. \cite{Izenov:L1-ERROR:sigmod-2024}, the extent of positional discrepancy between these vectors correlates with the likelihood of generating a suboptimal plan, with greater discrepancies indicating a higher probability of suboptimality, while no discrepancy suggests that the plan is likely to be optimal. Leveraging this insight, we carefully structure and featurize the query subplans to allow the classification model to effectively discern these positional discrepancies between ${\rho_k}$ and ${\hat{\rho}_k}$. By encoding the subplans into these vectors, the model can be trained to recognize patterns that are indicative of sub-optimal query plans, thereby enhancing its predictive accuracy in real-time query optimization scenarios.

\begin{figure}[htbp]
	\centering
	\includegraphics[width = .6\linewidth]{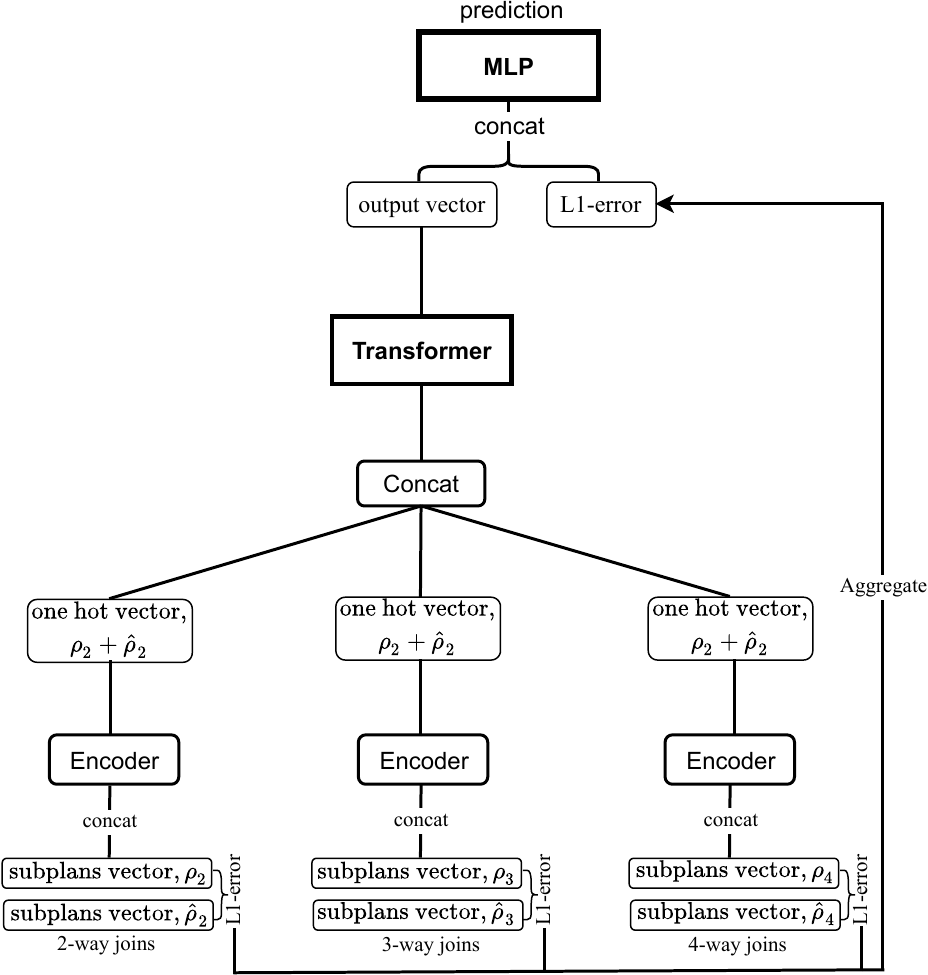}
	\caption{Classification model architecture.}
	\label{fig:classification_model}
\end{figure}

Subplans ${Q_{sub}}$ are categorized based on their join size, denoted by $k$, representing the number of tables involved in the join. The model is designed to handle joins of arbitrary size, ranging from $k=2$ up to any number of participating tables. For illustrative purposes, Figure \ref{fig:classification_model} depicts the process up to $k=4$. For each join size $k$, the positional vectors ${\rho_k}$ and ${\hat{\rho}_k}$ encapsulate the ordering of the corresponding subplans. This structured representation facilitates the model's ability to recognize and process the variations in subplan positions across different join sizes, ultimately contributing to its accurate assessment of plan sub-optimality.

\begin{figure*}[htbp]
	\centering
	\includegraphics[width = \linewidth]{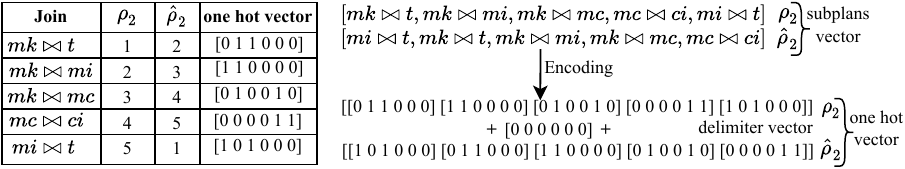}
	\caption{Subplan featurization as one-hot vector (first five 2-way joins from Figure \ref{fig:all_subs}).}
	\label{fig:featurization}
\end{figure*}

Initially, the subplan position vectors (${\rho}_{k}$ and ${\hat{\rho}}_{k}$) are fed into an encoder to generate a one-hot vector representation for each, as depicted in Figure \ref{fig:featurization}. Each join is represented by a unique one-hot vector with non-zero entries ranging from 2 to k, where each non-zero entry corresponds to a table involved in the join. A 2-way join is represented by two non-zero entries, each indicating a participating table. For instance, [0 1 0 0 0 0] represents table $mk$, and $t$ is defined by [0 0 1 0 0 0]. Join between $mk$ and $t$ --- $mk \Join t$ --- is represented by [0 1 1 0 0 0].

To ensure a clear distinction between the true cardinality-based position vector ${\rho}$ and the estimated cardinality-based position vector ${\hat{\rho}}$ within the model's input, we concatenate these vectors with a designated delimiter vector serving as a separator. This structured representation, formulated as $\rho$ + delimiter vector + $\hat{\rho}$, provides a clear demarcation, enabling the model to effectively differentiate and process the two distinct orderings of subplans.

\subsection{Model Architecture}\label{ssec:model}
The positional vectors $\rho$ and ${\hat{\rho}}$ represent two distinct orderings of the same set of subplans, ${Q_{sub}}$. Our primary objective is to discern the discrepancies between these sequences and predict whether the resultant query plan ${P}$, generated using the estimated cardinalities ${\hat{Y}}$, will be suboptimal (${P_{sopt}}$). This task inherently aligns with the framework of sequence classification, a technique widely employed in Natural Language Processing (NLP). To the best of our knowledge, the application of sequence classification to identify plan suboptimality, or within the broader domain of query optimization, remains unexplored.

Formulating this as a sequence classification problem presents several challenges. Primarily, the number of subplans ${N_k}$ varies for each join size $k$, leading to positional vectors ${\rho}_{k}$ and ${\hat{\rho}}_{k}$ of disparate lengths. To effectively learn the discrepancies between these sequences, it is imperative to adopt a model architecture capable of handling variable-length sequences and classifying the plan's sub-optimality based on these sequential disparities.

\subsubsection{Transformers}
The Transformer architecture \cite{Vaswani:transformers:NeurIPS}, renowned for its efficacy in handling variable-length sequences through self-attention mechanisms, is particularly well-suited for our task. The self-attention mechanism empowers the Transformer to evaluate the relevance of different segments within the input data, unburdened by the sequential order constraints that often limit traditional recurrent neural networks. This inherent flexibility renders Transformers exceptionally valuable for a diverse array of Natural Language Processing (NLP) tasks, notably including sequence classification.

However, the distinctive challenges posed by our application deviate from conventional NLP scenarios. In our context, each position within the sequence embodies a subplan, ${Q_{sub}}$, rather than a traditional lexical token. Despite this distinction, the fundamental challenge remains analogous to that in typical NLP applications: to effectively learn and comprehend the variances between two sequences, in our case, the true and estimated cardinality-based orderings of subplans, ${\rho}$ and ${\hat{\rho}}$ respectively. Consequently, we meticulously tailor the Transformer architecture to harmonize with the specific nuances of our problem domain. This adaptation primarily involves customizing the tokenization process, a critical step in preparing the input data for the Transformer's self-attention mechanisms.

\subsubsection{Customizing the Tokenization Process}
In contrast to traditional Natural Language Processing (NLP) applications where tokens typically represent words or subwords, PLANSEIVE's tokenization process is tailored to the unique structure and characteristics of query subplans, ${Q_{sub}}$. Each ``token'' in PLANSEIVE embodies a subplan, encapsulating its distinct attributes and operational context within the query plan $P$. To effectively process these specialized tokens, we modify the Transformer's tokenizer to recognize and segment the input data based on subplan characteristics. This customized tokenizer translates the one-hot encoded representations of subplans into unique token values using a predefined vocabulary, generated by considering all possible combinations of one-hot vectors.

The tokenizer operates as follows:
\begin{itemize}[leftmargin=1em, labelindent=1em, itemindent=0em, labelsep=0.5em]
    \item Subplan Encoding: For each subplan ${Q_{sub}}$, both the true cardinality-based position vector ${\rho}_{k}$ and the estimated cardinality-based position vector ${\hat{\rho}}_{k}$ are converted into one-hot encoded representations. These one-hot vectors are then mapped to their corresponding token values using the predefined vocabulary.
    
    \item Sequence Construction: The encoded subplans are then meticulously structured into sequences, commencing with a start-of-sequence token (\texttt{bos}), followed by the true subplan tokens, a separator token (\texttt{sep}), the estimated subplan tokens, and concluding with an end-of-sequence token (\texttt{eos}). This sequential arrangement enables the model to distinguish between the true and estimated subplan orderings within the input data.
    
    \item Padding and Attention Masking: To ensure compatibility with the Transformer model's input requirements, the sequences are adjusted to a uniform length by incorporating padding tokens. Additionally, an attention mask is generated to guide the model's self-attention mechanism, indicating which tokens within the sequence warrant attention.
    
    \item Tensor Conversion: Finally, the tokenized sequences and their corresponding attention masks are converted into tensors, the standard input format for the Transformer model. This transformation ensures that the model receives the input data in a structured and standardized manner, facilitating the effective application of self-attention mechanisms across the distinct subplans.
\end{itemize}

By meticulously encoding and structuring the subplans in this manner, the Transformer model in PLANSIEVE can leverage its self-attention mechanisms to gain a profound understanding of not only the individual subplans but also their relative positions and influences within the overall query plan. This sophisticated tokenization approach is integral to the model's ability to predict sub-optimality in query execution plans with high precision.

\subsubsection{Multilayer Perceptron (MLP)}
Subsequent to the sequence processing by the Transformer, the resultant output vector encapsulates the nuanced discrepancies between the positional vectors of subplans, ${\rho}_{k}$ and ${\hat{\rho}}_{k}$, across the various join sizes $k$ present in the query $Q$. This enriched vector representation serves as a crucial input for further analysis. When this representation is combined with the aggregated query-level L1-error, it forms a comprehensive feature set that encapsulates the potential suboptimality of the query plan $P$. The L1-error, as discussed in Section \ref{sec:prelim}, quantifies the cumulative deviation between the true and estimated cardinality-based orderings across all join sizes $k$, providing a nuanced measure of potential sub-optimality.

This concatenated feature vector, integrating both the Transformer’s output and the L1-error metrics, is then fed into a Multilayer Perceptron (MLP) for final classification. The MLP functions as the classification layer, tasked with interpreting the complex patterns and subtle differences encapsulated in the input vector. By leveraging the rich, structured information provided by the Transformer and L1-error calculations, the MLP makes a predictive assessment regarding the sub-optimality of the query plan $P$. This process is critical for determining whether the plan is likely to be sub-optimal ${P_{sopt}}$ or optimal ${P_{opt}}$.

The classification model, built on a GPT-2 architecture tailored for this specialized task, is configured to match the tokenizer’s vocabulary size and includes an embedding size and attention mechanism that are well-suited for query plan classification. By leveraging the structured information provided by the Transformer and the L1-error calculations, the MLP enables PLANSIEVE to make highly accurate predictions regarding the suboptimality of query plans. This integrated approach ensures that PLANSIEVE effectively identifies sub-optimal query execution plans.

\subsection{Model Implementation}
PLANSIEVE's classification model, implemented in PyTorch, employs a robust transformer-based architecture (GPT2 Sequence Classification) to capture the nuanced relationships between subplan positions within a query plan $P$. Input data, consisting of subplans ${Q_{sub}}$ and their respective true and estimated cardinalities ($Y$ and ${\hat{Y}}$), is preprocessed through a custom tokenizer that encodes these subplans into sequences.

The output from the transformer is then fed into a multilayer perceptron (MLP), which serves as the final classification layer. This MLP refines the model's predictions by incorporating additional features, such as the L1-error, which measures discrepancies between true and estimated cardinalities. The training process involves splitting the dataset into training and test sets, followed by extensive optimization using the AdamW optimizer. Regular evaluations on the test set ensure that the model maintains high accuracy and generalizes well to new data. Throughout the training, key metrics are tracked and visualized to monitor the model's performance, with the best-performing model saved for deployment. This implementation strategy ensures that PLANSIEVE's classification model is both effective and adaptable, capable of continuously improving its ability to predict suboptimal query plans as it encounters more queries.

\subsection{Training}\label{sec:training}

\subsubsection{Collecting Training Data} 
A fundamental challenge in learning-based approaches lies in the acquisition of a training dataset that adequately represents the diverse range of queries the system is likely to encounter. To address this, we adopt a strategy of randomly selecting $70\%$ of the queries from the workload to constitute our training set. This ensures that the model is exposed to a wide variety of query patterns, subplans, and cardinality distributions, fostering its ability to generalize effectively to unseen queries.

For each query $Q$ in the training set, we meticulously collect both the estimated cardinalities ${\hat{Y}}$ and the true cardinalities ${Y}$ of its constituent subplans ${Q_{sub}}$. The estimated cardinalities ${\hat{Y}}$ are obtained from the database's inherent cardinality estimator, while the true cardinalities ${Y}$ are derived from the actual execution of the corresponding subplans. This dual collection of cardinality data serves a pivotal role in the training process. By juxtaposing the estimated cardinalities ${\hat{Y}}$ with the actual cardinalities ${Y}$, the model can learn to discern and comprehend the discrepancies between these values, particularly in the context of the positional vectors ${\rho}$ (true cardinality-based) and ${\hat{\rho}}$ (estimated cardinality-based). This understanding is crucial for the model's ability to predict the likelihood of suboptimal query plans, as it forms the foundation for calculating the L1-error, a key metric in the model's classification capabilities.

\subsubsection{Offline Training Phase} 
The offline training phase is crucial for preparing PLANSIEVE's classification model to accurately identify suboptimal query plans. During this phase, the model undergoes extensive training on a representative dataset that includes queries, their enumerated subplans \({Q_{sub}}\), and both the true cardinalities $Y$ and estimated cardinalities ${\hat{Y}}$. The availability of true cardinalities --- obtained through the execution of subplans --- provides a unique advantage in the offline setting, allowing the model to deeply learn the relationship between ($\rho$, $\hat{\rho}$, L1-ERROR) and query plan sub-optimality.

The training process involves feeding the model with the structured representation of subplans, including their positional vectors $\rho$ and $\hat{\rho}$ (based on true and estimated cardinalities, respectively), along with the computed L1-error. The model, powered by a transformer-based architecture, learns to recognize patterns and discrepancies between these vectors, enabling it to predict the likelihood of suboptimality in query plans.

This comprehensive training, leveraging the representative dataset with true cardinalities, empowers the model to generalize and interpolate effectively between actual and estimated values. This capability is particularly critical in the online phase, where the model must operate with limited access to true cardinalities, relying primarily on a mixture of surrogate values ${Y_{s}}$ and learned true cardinalities $Y$ from historical query execution to make accurate predictions about the optimality of query plans.

\begin{figure}[htbp]
    \centering
    \includegraphics[width = 0.6\linewidth]{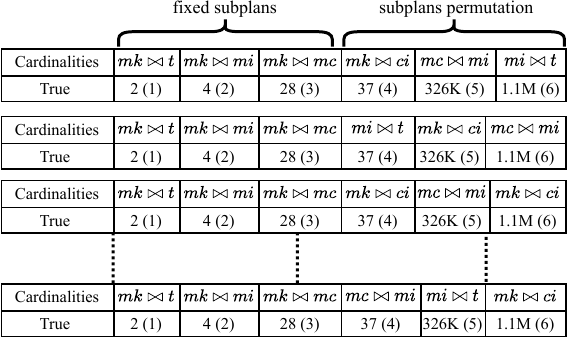}
    \caption{Subplan permutation strategy. Initial subplans are fixed while later subplans are permuted to highlight the importance of early subplan positioning.}
    \label{fig:subplans_permutation}
\end{figure}

\subsubsection{Not All Subplans Are Equally Important}
The subplans $Q_{sub}$ within a position vector $\rho$ or $\hat{\rho}$ are inherently ordered by their cardinalities (Figure \ref{fig:all_subs}), reflecting the optimizer's preference for subplans with lower cardinalities at each join size k. This implies a prioritization, where subplans appearing earlier in the vector are deemed more critical than those positioned later. The L1-error calculation inherently incorporates this prioritization by assigning higher penalties for incorrect placements of earlier subplans.

To ensure the model effectively learns this prioritization and recognizes the potential for suboptimal plans arising from misplacements of crucial early subplans, we augment the training data. Each training query $Q$ and its corresponding subplan position vectors are replicated multiple times. In each replica, the true cardinalities $Y$ of later subplans (those with higher positions in ${\rho}$) are randomly permuted, while the positions of the initial subplans remain fixed --- as shown in Figure \ref{fig:subplans_permutation}. The associated label (optimal or suboptimal) remains consistent across all replicas of a given query. This strategic modification of the training data emphasizes the importance of the relative ordering of early subplans, guiding the model to learn the critical impact of their positioning on overall plan optimality.

\subsection{Sub-Optimality Prediction}
During online operation, the classification model predicts the sub-optimality of a query plan $P$ without direct access to true cardinalities $Y$. It commences with true cardinality surrogate ${Y_{s}}$ provided by an external estimator, distinct from the DBMS's native one. As the system processes more queries, it incrementally refines its understanding of the actual cardinalities, as detailed in Section \ref{sec:learn_from_exec}.

For a given query $Q$, the prediction process begins by collecting all its subplans ${Q_{sub}}$ along with their corresponding cardinality estimates ${\hat{Y}}$. If available in the cache, true cardinalities $Y$ from historical execution data are used; otherwise, third-party estimates ${Y_{s}}$ are employed. Subsequently, the L1-error is computed for the query.

The subplans are then transformed into a feature representation, as elaborated in Section \ref{ssec:query_representation}. These feature vectors are processed by a transformer model, which learns the relative order differences among the subplans. The output vector from the transformer, combined with the computed L1-error, is then input into a classification layer. This layer predicts the likelihood that the given plan P, based on the estimates, is sub-optimal or not.

\subsection{Workload and Data Changes}\label{sec:change}

\subsubsection{Adapting to Workload Changes} 
Changes in workload characteristics can lead to cache misses for selection-sensitive patterns, as these patterns necessitate an exact match of all selection conditions present in the pattern to correspond with a sub-query ${Q_{sub}}$. In contrast, patterns constructed using Approaches 1 and 2, as discussed in Section \ref{ssec:cache-hit}, exhibit greater flexibility with respect to selection conditions. Specifically, Approach 1 generates patterns entirely devoid of selection criteria, while Approach 2 merely notes the presence of selection conditions on specific tables, without capturing their precise details. 

Both approaches demonstrate adaptability to shifts in workload patterns. They can be readily modified to maintain a running average of cardinalities, prioritizing recent workloads to reflect the current query distribution. The only scenario where these approaches encounter a mismatch is when the new workload introduces novel join conditions ${J}$ between tables, necessitating the creation of new patterns and the accumulation of fresh cardinality data. 

\subsubsection{Data Changes}
Data modifications, encompassing updates, deletions, or varied data ingestion patterns, can significantly impact the accuracy of cardinality estimation. Both default system estimators and third-party tools rely on up-to-date statistics to reflect the true data distribution. As the underlying data undergoes changes, previously collected statistics may become stale, leading to discrepancies between estimated cardinalities ${\hat{Y}}$ and the actual cardinalities ${Y}$.

To mitigate this, database systems like PostgreSQL employ mechanisms to periodically refresh their statistical data. The \texttt{ANALYZE} command is instrumental in this process, enabling the system to gather updated statistics and recalibrate its cardinality estimates to better align with the current data state.

Machine learning-based models for cardinality estimation can be broadly categorized into two types: those trained on static data snapshots and those trained on dynamic workload patterns. Models in the former category, while potentially offering high accuracy on the training data, may exhibit reduced adaptability to changes in the underlying data distribution. In contrast, models trained on workload patterns are inherently designed to adapt to evolving query trends. 

Irrespective of the training approach, these models necessitate periodic retraining to maintain their estimation accuracy in the face of data and workload fluctuations. This retraining is crucial for incorporating recent modifications into the models, ensuring the continued precision of their predictions.

In this study, we leverage the default data adaptation strategies embedded within each model. While our primary focus is not on refining these adaptation processes, we acknowledge the importance of maintaining model currency with respect to dataset changes to ensure the validity and reliability of our experimental results.

\section{EXPERIMENTAL EVALUATION}\label{sec:experiments}
We conduct a comprehensive experimental study to rigorously evaluate the PLANSIEVE framework across various real-world scenarios. The study commences with an offline training phase, where the PLANSIEVE classification model (PLANSIEVE's CM) is trained on a dataset meticulously curated to encompass a diverse range of queries $Q$. This offline phase establishes a foundation for the model's subsequent performance in the online phase, where true cardinalities $Y$ are initially unavailable, challenging the model's ability to accurately identify sub-optimal plans ${P_{sopt}}$.

In our evaluations, we adopt a threshold $c=1$ for defining plan sub-optimality, which implies that a plan is considered sub-optimal if it costs more than, the optimal plan calculated using accurate cardinalities.

Our evaluation begins by assessing PLANSIEVE's capability to operate effectively in this challenging online scenario, where it must rely on surrogate cardinality estimates ${Y_{s}}$. As the experiments progress, we gradually introduce true cardinality values $Y$ that have been cached from prior executions. This allows us to observe how the model adapts to a mixed environment of true $Y$ and surrogate cardinalities ${Y_{s}}$. The investigation focuses on several key questions:

\begin{itemize}[leftmargin=*,noitemsep,nolistsep]
	\item How does PLANSIEVE perform offline, particularly compared to other classification models?

	\item How does PLANSIEVE handle the unavailability of true cardinalities ($Y$) during online operation?
	
	\item How does learning from historical query execution data impact PLANSIEVE's ability to classify sub-optimal plans?
	
	\item What are the computational overheads of training and prediction with PLANSIEVE's CM?
	
	\item How does the choice of third-party cardinality estimator impact PLANSIEVE's online performance?
	
	\item How does PLANSIEVE perform on workloads like JOB-LIGHT-RANGES, which differ from its training data?
\end{itemize}

These questions guide our experimental analysis, aiming to provide a deep understanding of PLANSIEVE's potential for practical deployment in dynamic query processing environments. Through this rigorous evaluation, we seek to uncover not only the strengths of the PLANSIEVE framework but also any limitations, offering insights into its applicability and effectiveness in real-world database systems. The artifacts required to replicate the experiments are available online~\cite{plansieve-code}.

\subsection{Model Implementation Details}
The PLANSIEVE classification model (PLANSIEVE's CM), implemented in PyTorch, leverages a transformer-based architecture (specifically, GPT2ForSequenceClassification) to capture the intricate relationships between subplans ${Q_{sub}}$ within a query plan $P$. The model's configuration is adaptable to the dataset's characteristics, with the transformer architecture tailored to handle varying sequence lengths and complexities. For instance, in the STATS-CEB-SCALE workload, the transformer is configured with 6 layers, 8 attention heads, and an embedding dimension of 128, processing sequences of up to 27 tokens. In contrast, for the JOB-LIGHT-SCALE workload, the configuration is adjusted to accommodate shorter sequences with an embedding dimension of 64 and up to 23 tokens.

The transformer's output, specifically the final hidden state, serves as a compact representation of the subplans relative order. This representation, along with the query-level L1-error, is fed into a Multilayer Perceptron (MLP) for final classification. The MLP, typically with a hidden layer and a 2-neuron output layer with softmax activation, is optimized for binary classification tasks, predicting the likelihood of a query plan P being suboptimal.

Model training employs the AdamW optimizer with a OneCycleLR scheduler, and hyperparameters like learning rate, number of epochs, and batch size are fine-tuned based on the dataset's complexity. The CrossEntropyLoss function is used, and accuracy serves as the primary evaluation metric. The best-performing model is checkpointed for deployment. This adaptable implementation enables PlanSieve to provide robust and precise sub-optimality predictions across diverse query processing scenarios.

\subsection{Workload}\label{ssec:workload}
This subsection details the benchmarks used to evaluate PLANSIEVE. The choice of benchmark is influenced by the fact that most open-source third-party cardinality estimators, including the one used as a surrogate estimator in this work, are machine learning-based \cite{Hilprecht:DeepDB:pvldb-2020, Kipf:LCECJDL:cidr-2019, Yang:NeuroCard:pvldb-2021, Kipf:DeepSketches:SIGMOD-2019, Liu:CEUNN:cascon-2015,Malik:BBAQCE:cidr-2007,Ortiz:EADLCE:arxiv-2019,Woltmann:CEL:aiDM-2019} and may not have strong support for queries with LIKE predicates or string selections. PLANSIEVE can work with any benchmark if a suitable third-party cardinality estimator is available. We follow prior work \cite{Han:CEB:2021, Kipf:LCECJDL:cidr-2019, Yang:NeuroCard:pvldb-2021} and utilize JOB-LIGHT and STATS-CEB for our evaluation. These benchmarks are widely adopted for evaluating machine learning-based components in database query optimization due to their diverse query complexity and the challenges they pose for cardinality estimation.

\subsubsection{JOB-LIGHT}  
JOB-LIGHT, a modified version of the Join Order Benchmark (JOB) \cite{JOB-github}, is specifically designed to assess machine learning-based cardinality estimators \cite{Kipf:LCECJDL:cidr-2019}. This benchmark consists of 70 queries that involve a range of join operations across six different tables. Importantly, JOB-LIGHT deliberately excludes queries with predicates on strings and disjunctions, focusing instead on queries with between one and four joins. The queries primarily utilize equality predicates on attributes from dimension tables and are structured predominantly in star join schemas centered around a single fact table. This design aligns with the constraints of many machine learning-based cardinality estimators, which often perform better in the absence of complex string predicates.

\subsubsection{JOB-LIGHT-ranges} 
JOB-light-ranges \cite{Yang:NeuroCard:pvldb-2021} is an extension of the JOB-light benchmark, consisting of 1000 queries that follow similar patterns to those in JOB-light but with a key distinction. Unlike JOB-light, which features range predicates on just one attribute, JOB-light-ranges includes range predicates on multiple attributes. This enhancement allows for a more complex and comprehensive testing environment, simulating a broader array of real-world scenarios where queries may involve conditions on several fields simultaneously.

\subsubsection{STATS-CEB}
The STATS-CEB benchmark \cite{Han:CEB:2021}, integrates the STATS dataset from the Stats Stack Exchange network with a curated query workload. This benchmark aims to represent real-world data environments, encompassing a diverse set of multi-table join queries without the simplifications often seen in other benchmarks. With 146 queries spanning a range of complexities, including various join types and extensive filter conditions, STATS-CEB provides a robust platform for assessing the efficacy of Machine Learning-based methods. Its comprehensive design, featuring complex data distributions and a rich join schema, more accurately simulates the challenges encountered in practical database systems compared to earlier benchmarks.

\subsubsection{Limitations of Existing Benchmarks} 
While JOB-LIGHT and STATS-CEB offer a variety of query scenarios, their limited number of queries presents challenges for robust classification model training and evaluation in PLANSIEVE. This limitation is further compounded by the class imbalance inherent in these benchmarks, which often yield only a small number of sub-optimal plans relative to the total query count. To overcome these constraints, we have developed a flexible and scalable framework for benchmark augmentation. This framework allows us to systematically expand the existing benchmarks, thereby increasing the diversity and quantity of queries available for training and evaluation, and mitigating the effects of class imbalance.

\begin{figure*}[htbp]
	\centering
	\includegraphics[width=.8\textwidth]{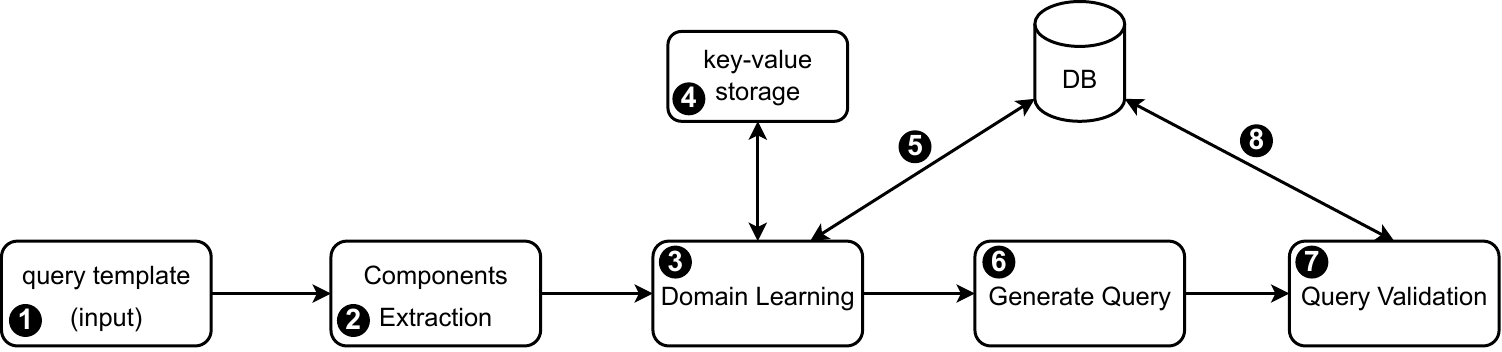}
	\caption{Workload scaling framework.}
	\label{fig:scaling_framework}
\end{figure*}

\textbf{Workload Scaling Framework.}
The framework, as depicted in Figure \ref{fig:scaling_framework}, facilitates the systematic generation of a diverse and extensive array of queries from existing benchmark workloads, such as JOB-LIGHT \cite{Kipf:LCECJDL:cidr-2019} and STATS-CEB \cite{Han:CEB:2021}.

The augmentation process begins with the Component Extraction module \circled{2}, which analyzes a given template query -- existing queries from workload -- \circled{1} to identify its key elements---such as selection predicates, join conditions, and involved tables. This step deconstructs the query's structure, laying the groundwork for generating new queries that are both syntactically and semantically valid.

Following this, the Domain Learning phase \circled{3} determines the range of permissible values for each attribute referenced in the query predicates. The module first checks a key-value storage \circled{4} to ascertain whether the domain information for a given column has been previously gathered. If the domain remains unexplored, the module formulates targeted queries to extract the necessary domain data from the database \circled{5}, subsequently storing this information for future use. This process ensures that the generated queries align with the actual data distribution, maintaining their relevance and fidelity to the underlying dataset.

In the subsequent Query Generation phase \circled{6}, this domain knowledge is employed to create new queries by adjusting selection predicates and, where needed, modifying join conditions. These adjustments adhere to the structural framework of the original template query, retaining its core characteristics while introducing controlled variations. Each newly generated query undergoes a thorough Query Validation process \circled{7}, verifying both its syntactic correctness and semantic validity to ensure that it yields meaningful and non-empty results.

This proposed framework enhances the evaluation process, particularly when the number of challenging benchmark queries is limited. In the JOB-LIGHT benchmark, for example, only 2 out of 70 queries resulted in sub-optimal plans, with the remaining 68 yielding optimal plans. This pronounced imbalance severely restricts the effectiveness of model training and evaluation. Our framework tackles this limitation by systematically augmenting the workload in two ways: first, by scaling existing queries while maintaining their fundamental join structure; and second, by introducing novel or modified selection and join predicates. This systematic diversification substantially expands the available query landscape. Specifically, we generated 1,357 new queries for JOB-LIGHT and 2,380 for STATS-CEB, culminating in the expanded JOB-LIGHT-SCALE and STATS-CEB-SCALE workloads. These augmented workloads furnish a richer and more balanced dataset, thereby facilitating more robust evaluation and development of database systems.

\subsection{Offline Phase}\label{ssec:offline}
\begin{figure*}[htbp]
    \centering
    \includegraphics[width=\textwidth]{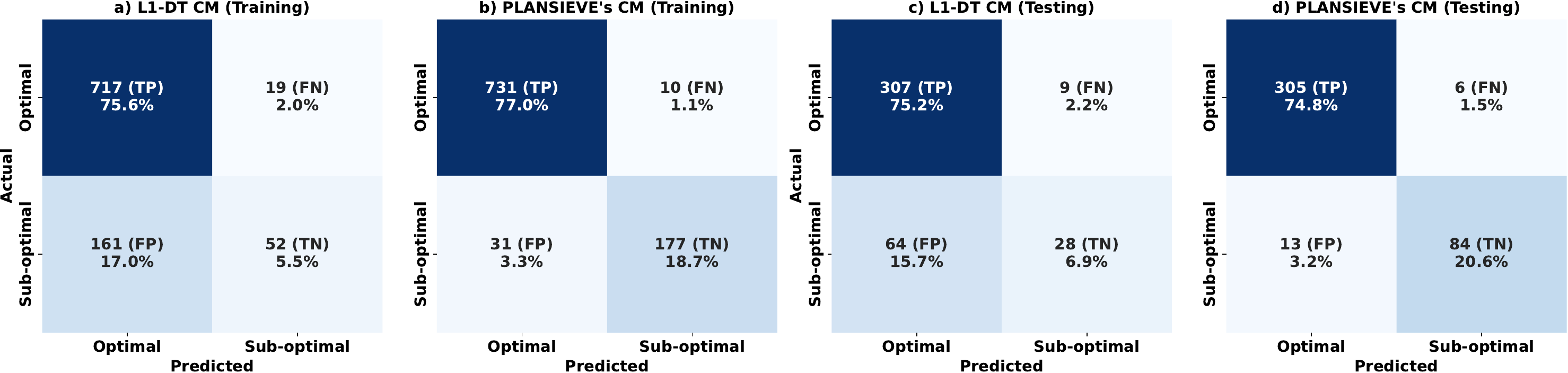}
    \caption{Confusion matrix for the offline phase (JOB-LIGHT-SCALE).}
    \label{fig:offline_job_light}
\end{figure*}

\begin{figure*}[htbp]
    \centering
    \includegraphics[width=0.95\textwidth]{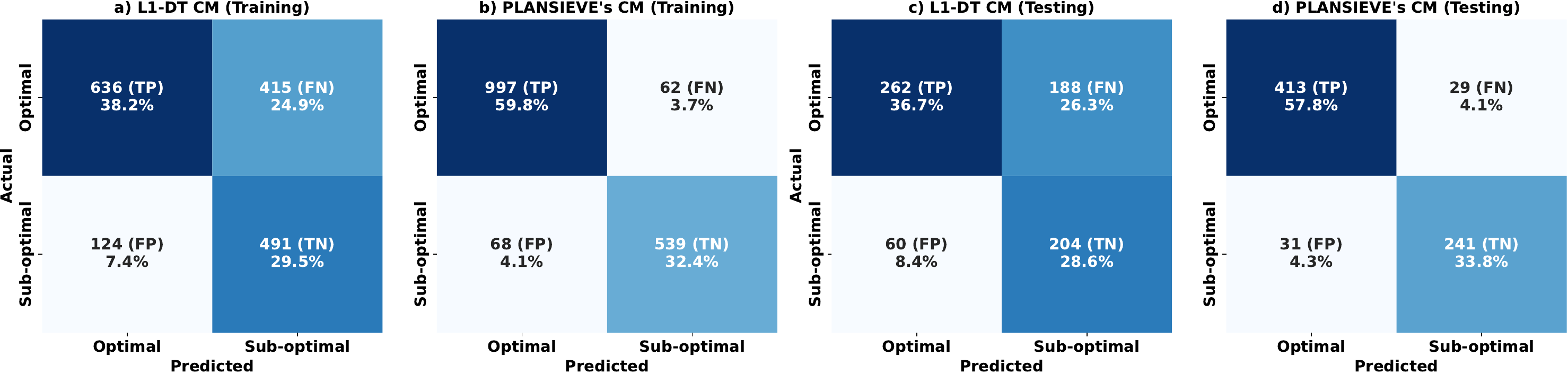}
    \caption{Confusion matrix for the offline phase (STATS-CEB-SCALE).}
    \label{fig:offline_stats_ceb}
\end{figure*}

PLANSIEVE's classification model (PLANSIEVE's CM) is trained entirely offline to identify suboptimal query plans, requiring no online retraining. This training is performed on the JOB-LIGHT-SCALE and STATS-CEB-SCALE datasets, which comprise queries, their enumerated subplans, and corresponding true and estimated cardinalities.  The offline phase focuses on learning the relationship between input features and query plan suboptimality.  Datasets are partitioned into 70\% for training and 30\% for testing, with model performance evaluated using confusion matrices (Figure \ref{fig:offline_job_light}, \ref{fig:offline_stats_ceb}). This offline training enables PLANSIEVE's CM to effectively identify suboptimal plans during the online phase, where true cardinalities are unavailable.

To evaluate the performance of PLANSIEVE's CM, we utilize confusion matrices to provide a detailed breakdown of its predictive capabilities. These matrices categorize predictions into four distinct outcomes: True Positives (TP), True Negatives (TN), False Positives (FP), and False Negatives (FN).  Given our focus on accurately identifying sub-optimal plans (TN), we prioritize minimizing the misclassification of such plans (FP).  For both training and testing phases, we report the number and percentage of queries falling into each category, enabling a thorough assessment of the model's accuracy and effectiveness in identifying sub-optimal query plans.

To further validate the effectiveness of PLANSIEVE's CM, we benchmark it against the decision tree-based classification model (L1-DT CM) proposed by Izenov et al. \cite{Izenov:L1-ERROR:sigmod-2024}. This comparison highlights the differences in input features and architectural complexities between the two models.  PLANSIEVE's CM utilizes a sophisticated machine learning framework that incorporates both the relative order of subplans and L1-ERROR, enabling a more nuanced classification approach. In contrast, L1-DT CM employs a simpler decision tree model that relies solely on L1-ERROR as its input feature. This simpler design, with parameters optimized through grid search and cross-validation, favors directness over the more complex, multi-feature approach of PLANSIEVE's CM.

\begin{table*}[htbp]
	\centering
	\small
	\begin{tabularx}{\textwidth}{l l r r r r}
	\toprule
	\textbf{Dataset} & \textbf{Model} & \textbf{Training Queries} & \textbf{Testing Queries} & \textbf{Training Time} & \textbf{Prediction Time} \\
	\midrule
	\multirow{2}{*}{JOB-LIGHT-SCALE} 
	 & PLANSIEVE’s CM & \multirow{2}{*}{1,357} & \multirow{2}{*}{407} &  232.72 s & 1.32 ms \\ 
	 & L1-DT CM       &  &  & 4.2 ms & 2.57 $\mu$s \\
	\midrule
	\multirow{2}{*}{STATS-CEB-SCALE} 
	 & PLANSIEVE's CM & \multirow{2}{*}{1,667} & \multirow{2}{*}{715} & 518.66 s & 1.03 ms \\
	 & L1-DT CM       &  &  & 37 s & 1.26 $\mu$s \\
	\bottomrule       
	\end{tabularx}
	\caption{Computational overhead: PLANSIEVE's CM vs.\ L1-DT CM.}
	\label{tab:computational_overhead}
\end{table*}

\textbf{Computational Overhead.}
Table \ref{tab:computational_overhead} presents the computational overhead associated with PLANSIEVE's CM and L1-DT CM, quantifying training time, total prediction time, and average prediction time per query across the JOB-LIGHT-SCALE and STATS-CEB-SCALE workloads. Notably, the transformer-based architecture of PLANSIEVE's CM entails more extended training periods relative to the more streamlined L1-DT CM. Despite its higher computational cost, PLANSIEVE's CM delivers enhanced accuracy, as demonstrated in prior analyses. This reflects a fundamental trade-off between computational efficiency and model sophistication, underscored by PLANSIEVE's CM's advanced feature representation and intricate architectural framework.

Regarding storage requirements, PLANSIEVE's CM demands significantly larger model sizes compared to L1-DT CM due to its complex architecture—approximately 3.72 MB and 9.59 MB for the JOB-LIGHT-SCALE and STATS-CEB-SCALE models, respectively. Importantly, the offline nature of the training process does not impact PLANSIEVE's performance during online query processing. While PLANSIEVE's current per-query prediction time is higher than that of L1-DT CM, it remains practical for many real-time applications. Future research focusing on model architecture optimization and implementation enhancements is anticipated to further reduce this overhead, enhancing PLANSIEVE's viability for performance-critical scenarios.

\subsubsection{JOB-LIGHT-SCALE Workload} 
The evaluation of the JOB-LIGHT-SCALE workload underscores the efficacy of PLANSIEVE's classification model (PLANSIEVE's CM) in both the training and testing phases. During training, PLANSIEVE's CM correctly identified 731 out of 949 queries as optimal (true positives) and 177 as suboptimal (true negatives). The model exhibited a low error rate, with 10 false positives --- suboptimal plans incorrectly classified as optimal --- and 31 false negatives --- optimal plans misclassified as suboptimal, which together constitute 4.4\% of the total training queries.

In contrast, the L1-DT CM model, while identifying a comparable number of true positives (717), struggles in correctly identifying suboptimal plans, evidenced by a notably lower true negative count (52). Furthermore, it had a higher misclassification rate --- 19\% --- with 161 false positives and 19 false negatives.

In the testing phase, PLANSIEVE's CM maintained its strong performance, correctly classifying 305 out of 408 queries as optimal and 84 as suboptimal. The model's error rate remained low at 4.7\%, with 13 false positives and 6 false negatives. L1-DT CM, on the other hand, exhibited similar challenges as in the training phase, particularly in identifying suboptimal plans. It achieved 307 true positives and only 28 true negatives. The model also had a higher misclassification rate --- 17.9\% --- with 64 false positives and 9 false negatives.

The comparative analysis reveals that while both models demonstrate reasonable accuracy in identifying optimal plans, PLANSIEVE's CM shows a distinct advantage in correctly classifying sub-optimal plans. This is evident in its lower false positive rate across both the training and testing phases, indicating a reduced likelihood of misclassifying sub-optimal plans as optimal. Such misclassifications can have detrimental consequences in real-world database systems, as they can lead to the execution of inefficient query plans, resulting in increased query response times and resource wastage. PLANSIEVE's CM's ability to minimize these errors highlights its potential for enhancing query optimization effectiveness in the online phase.

\subsubsection{STATS-CEB-SCALE Workload}

Transitioning to the more intricate and diverse STATS-CEB-SCALE workload, the evaluation further reinforces the consistent and robust nature of PLANSIEVE's classification model (PLANSIEVE's CM). During the training phase, PLANSIEVE's CM correctly classified 997 queries as optimal (true positives, TP) and 539 as suboptimal (true negatives, TN), with a relatively low error rate --- 7.8\% of the total training queries --- resulting in 68 false positives (FP) and 62 false negatives (FN). In contrast, the L1-DT CM model, while identifying 636 true positives and 491 true negatives, exhibited a notably higher misclassification rate of 32.3\%, leading to 415 false negatives --- optimal plans identified as suboptimal, accounting for 24.9\% of total queries --- and 124 false positives. Notably, this misclassification pattern differs from the JOB-LIGHT-SCALE workload, where the predominant errors were false positives. This variation suggests that PLANSIEVE's CM incorporation of features such as the relative order of subplans enhances its ability to detect subtle patterns indicative of suboptimal plans, leading to more accurate classifications.

During the testing phase, PLANSIEVE's CM continued to demonstrate its effectiveness, correctly classifying 413 true positives and 241 true negatives while maintaining a minimal error rate with 31 false positives and 29 false negatives. On the other hand, the L1-DT CM model struggled to accurately identify optimal plans, resulting in 188 false negatives and 60 false positives, reflecting similar trends observed during training. Despite securing 262 true positives and 204 true negatives, L1-DT CM's performance discrepancies further highlight the advantages of PLANSIEVE's CM comprehensive approach in addressing the complexities and subtleties of the STATS-CEB-SCALE workload.

This comparative analysis across both training and testing phases underscores PLANSIEVE-CM's consistent ability to accurately classify both optimal and sub-optimal plans. While the L1-DT CM model demonstrates reasonable accuracy in identifying suboptimal plans, it shows a higher susceptibility to errors --- particularly in producing a higher number of false negatives, where optimal plans are incorrectly identified as suboptimal --- highlights its limitations. PLANSIEVE-CM's robustness in reducing such misclassifications positions it as a more reliable tool for optimizing query performance in complex, dynamic environments.

\subsubsection{Summary} 

The comprehensive offline evaluation conducted on both the JOB-LIGHT-SCALE and STATS-CEB-SCALE workloads underscores the robustness and adaptability of PLANSEIVE's classification model (PLANSEIVE's CM). In these diverse query environments, PLANSEIVE's CM consistently outperformed the baseline L1-DT CM model, particularly in minimizing critical misclassifications. In the JOB-LIGHT-SCALE workload, PLANSEIVE's CM demonstrated a notable reduction in errors, especially in avoiding the misclassification of suboptimal plans as optimal. The evaluation on the STATS-CEB-SCALE workload further validated PLANSEIVE's CM's strengths, as the model continued to maintain low error rates and effectively managed both false positives and false negatives. In contrast, the L1-DT CM model exhibited higher misclassification rates, particularly false negatives, highlighting its limitations.

While PLANSEIVE's CM incurs higher computational overhead compared to L1-DT CM, this trade-off is justified by its superior capability to prevent costly misclassifications, particularly in complex and dynamic query environments. L1-DT CM, originally designed for post-analysis --- offline evaluation, demonstrates inconsistent performance across different workloads due to its simplistic nature, which underscores the need for a more robust solution like PLANSEIVE's CM. Despite higher computational demands, PLANSEIVE's CM offers a more reliable and accurate classification. Since the model is trained entirely offline, the training time does not impact its real-time performance, and although prediction times are higher compared to L1-DT CM, they remain within practical limits for real-world applications. Given these factors, PLANSEIVE's framework opts for PLANSEIVE's CM to ensure consistent and accurate results, providing a solid foundation for the online evaluation, where the model's performance will be tested under dynamic, real-time conditions.

\subsection{Online Phase}\label{ssec:online_eval}

The online evaluation of PLANSIEVE's CM, utilizing the JOB-LIGHT-SCALE and STATS-CEB-SCALE datasets, investigates its performance under varying conditions of cardinality information availability. We hypothesize that the model's ability to identify suboptimal plans will progressively improve as it transitions from relying solely on surrogate true cardinalities ($Y_s$) to incorporating increasing amounts of true cardinalities ($Y$) obtained during query execution.

While the offline training phase provides the model with access to true cardinalities for all subplans, the online setting initially relies on surrogate true cardinalities provided by a third-party estimator (DeepDB \cite{Hilprecht:DeepDB:pvldb-2020}). As PLANSIEVE processes queries, these initial estimates are refined by progressively integrating observed true cardinalities.

To systematically evaluate this adaptive behavior, we define five scenarios simulating different combinations of $Y_s$ and $Y$. These range from utilizing solely $Y_s$ to a gradual incorporation of $Y$ in increments of 25\%, 50\%, and 75\%, culminating in a scenario where only $Y$ is employed. Performance is analyzed through confusion matrices, categorizing predictions into true positives (TP), false positives (FP), true negatives (TN), and false negatives (FN). This analysis focuses on assessing the model's ability to effectively identify suboptimal plans and minimize their misclassification as more true cardinality information becomes available.

\subsubsection{JOB-LIGHT-SCALE Workload}
\begin{figure*}[htbp]
    \centering
    \includegraphics[width=\textwidth]{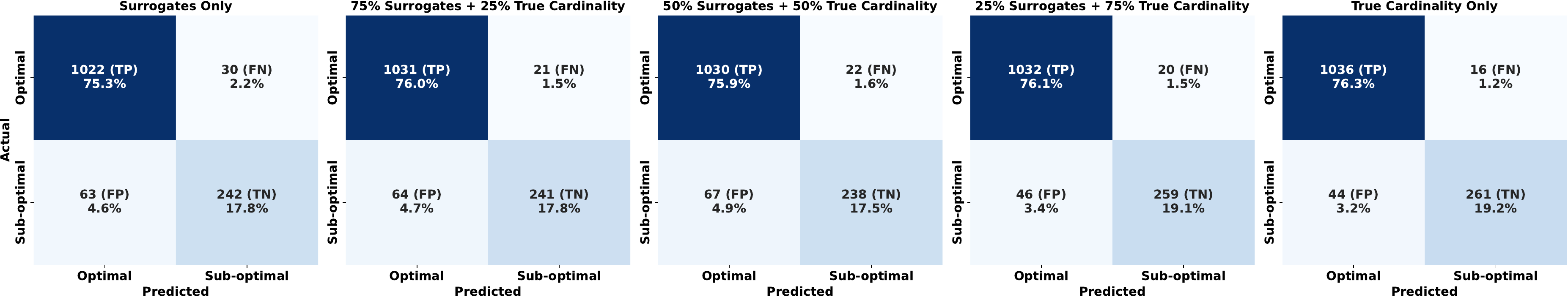}
    \caption{PLANSIEVE's online performance on JOB-LIGHT-SCALE under varying cardinality information (${Y_{s}}$ + ${Y}$).}
    \label{fig:cm_job_light}
\end{figure*}

The JOB-LIGHT-SCALE workload serves as a comprehensive testbed for evaluating the online adaptability of the PLANSIEVE classification model (PLANSIEVE's CM) under varying cardinality scenarios. The model's performance was rigorously assessed across five distinct scenarios, ranging from complete reliance on surrogate cardinalities ${Y_{s}}$ to exclusive utilization of true cardinalities $Y$, with intermediate stages incorporating a blend of both. The confusion matrices depicted in Figure \ref{fig:cm_job_light} provide a detailed breakdown of the model's performance across these scenarios, showcasing its ability to progressively refine its predictions and enhance accuracy as it processes more queries and incorporates a greater proportion of true cardinality data.

\textbf{Surrogates Only Scenario (100\% ${Y_{s}}$).}
Initially, the model's performance was evaluated relying solely on the surrogate cardinalities $Y_{s}$ provided by DeepDB. Under these conditions, the model achieved 1,022 true positives (TP) and 242 true negatives (TN), leading to an overall classification accuracy of 93.15\%. Despite this strong performance, the model did encounter 63 false positives (FP) and 30 false negatives (FN), which resulted in a misclassification rate of 6.8\%. Given that the model was entirely dependent on surrogate cardinalities $Y_{s}$, this misclassification rate is relatively low. This outcome suggests that the estimates provided by DeepDB are generally sufficient to preserve the relative order of subplans and minimize the L1-ERROR, thereby maintaining the model's effectiveness even in the absence of true cardinality data $Y$.

\textbf{Integration of True Cardinalities.}
As true cardinalities ${Y}$ were incrementally integrated, starting with a 75\% surrogates ${Y_{s}}$ and 25\% true cardinality ${Y}$ mix, the model showed slight improvements and fluctuations:

\vspace{0.5em}
\vspace{-\topsep}
\begin{itemize}[leftmargin=1em, labelindent=1em, itemindent=0em, labelsep=0.5em]
	\item 75\% Surrogates + 25\% True Cardinalities (75\% ${Y_{s}}$ + 25\% ${Y}$): At this setting, we observed a slight improvement in the model's performance. The number of true positives (TP) increased to 1031, representing 76.0\% of the total queries, indicating a marginal gain in the identification of optimal plans. The false negative rate (FN) decreased to 1.5\%, suggesting improved detection of optimal plans. However, both the false positive rate (FP) and true negative rate (TN) remained relatively stable at 4.7\% and 17.8\%, respectively. This configuration led to a marginal increase in overall classification accuracy to 93.8\%, compared to the surrogates-only scenario, demonstrating that even a small integration of true cardinalities can positively impact query plan classification.

	\item 50\% Surrogates + 50\% True Cardinalities (50\% ${Y_{s}}$ + 50\% ${Y}$): With an equal mix of surrogates ${Y_{s}}$ and true cardinalities ${Y}$, the model's performance exhibits a slight adjustment. The true positives (TP) marginally decrease to 1030 (75.9\%), and true negatives (TN) slightly reduce to 238 (17.5\%). Simultaneously, there is a small increase in false negatives (FN) to 22 (1.6\%) and false positives (FP) to 67 (4.9\%). This configuration leads to a minor dip in overall classification accuracy to 93.38\%. This outcome suggests that while incorporating true cardinalities typically enhances the model's accuracy, the equal mix of surrogate and true data introduces a challenge. The model likely faces difficulty in reconciling the potential inconsistencies between surrogate estimates ${Y_{s}}$ and true cardinalities ${Y}$, which may result in a slight reduction in its ability to correctly classify plans.

	\item 25\% Surrogates + 75\% True Cardinalities (25\% ${Y_{s}}$ + 75\% ${Y}$): As the proportion of true cardinalities ${Y}$ increases to 75\%, the model shows a marked improvement in performance. True positives (TP) rise to 1032 (76.1\%), and true negatives (TN) also increase to 259 (19.1\%). This improvement is accompanied by a reduction in false negatives (FN) to 20 (1.5\%) and a more substantial drop in false positives (FP) to 46 (3.4\%). The overall classification accuracy improves to 95.14\%. This configuration highlights the model's enhanced ability to accurately classify both optimal and suboptimal plans as more true cardinality data is integrated, reflecting the model's growing confidence and precision as the PLANSIEVE framework refines more surrogate estimates ${Y_{s}}$ with true cardinalities ${Y}$ obtained from query execution.
\end{itemize}
\vspace{-\topsep}
\vspace{0.5em}

\textbf{True Cardinalities Only (100\% ${Y}$).}
Finally, when the model operates exclusively on true cardinalities ${Y}$, it achieves the most favorable performance, as expected. The number of true positives further increases to 1036 (76.3\%), and true negatives rise to 261 (19.2\%). Concurrently, we observe a reduction in both false negatives to 16 (1.2\%) and false positives to 44 (3.2\%). The overall classification accuracy reaches its peak at 95.58\%. This configuration underscores the model's ability to attain optimal performance when provided with the most accurate cardinality information, highlighting the critical role of true cardinalities in refining the model's predictive capabilities. As PLANSIEVE progressively incorporates true cardinalities gleaned from query executions into its cache, the model's reliance on surrogates ${Y_{s}}$ diminishes. This enables the model to leverage increasingly precise cardinality information, leading to enhanced discrimination between optimal and suboptimal plans and, consequently, improved classification accuracy.

\begin{figure*}[htbp]
	\centering
	\includegraphics[width = \textwidth]{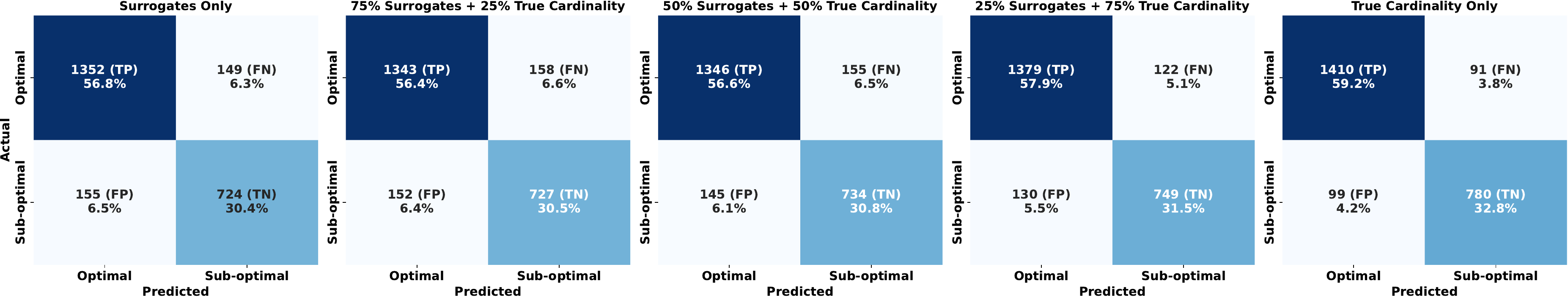}
	\caption{PLANSIEVE's online performance on STATS-CEB-SCALE under varying cardinality information (${Y_{s}}$ + ${Y}$).}
	\label{fig:cm_stats_ceb}
\end{figure*}

\subsubsection{STATS-CEB-SCALE Workload}
Building on the insights gained from the JOB-LIGHT-SCALE workload evaluation, we extend our analysis to assess the performance of PLANSIEVE's classification model under the STATS-CEB-SCALE workload. This workload introduces a unique and challenging context, allowing us to evaluate the model's robustness and precision across a more diverse set of queries, each characterized by varied cardinality conditions. Consistent with our approach in the JOB-LIGHT-SCALE workload, the model is rigorously tested across a spectrum of scenarios, ranging from the exclusive use of surrogate cardinalities ${Y_{s}}$ to scenarios relying solely on true cardinalities ${Y}$, with intermediate configurations combining both types of data. The outcomes of these evaluations, as captured in the confusion matrices presented in Figure \ref{fig:cm_stats_ceb}, provide a comprehensive view of the model's adaptability and accuracy in handling the complexities of the STATS-CEB-SCALE workload in real-time.

\noindent
\textbf{Surrogates Only Scenario (100\% ${Y_{s}}$).} Initially operating solely with surrogate cardinalities ${Y_{s}}$, the model achieves 1352 true positives (TP) for the Optimal class, representing 56.8\% of the queries, and 724 true negatives (TN) for the Sub-optimal class, accounting for 30.4\% of the queries. However, the model also encounters 149 false negatives (FN) (6.3\%) and 155 false positives (FP) (6.5\%), leading to an overall classification accuracy of 87.2\%. This misclassification rate is notably higher than in the comparable scenario for the JOB-LIGHT-SCALE workload, reflecting the increased complexity and variability within the STATS-CEB-SCALE workload. Despite these challenges, the model demonstrates a commendable baseline performance, indicating that even with surrogate cardinalities ${Y_{s}}$, PLANSIEVE's CM can leverage its offline training to effectively differentiate between optimal and sub-optimal plans to a certain extent. It is important to note that as PLANSIEVE processes more queries and accumulates true cardinalities in its cache, we anticipate a progressive refinement of its predictive capabilities, showcasing the framework's ability to learn from query execution data and adapt to the evolving cardinality landscape.

\noindent
\textbf{Integration of True Cardinalities.} The strategic incorporation of true cardinalities ${Y}$ progressively enhanced the model's accuracy in distinguishing between optimal and sub-optimal classifications under the STATS-CEB workload:

\begin{itemize}[leftmargin=1em, labelindent=1em, itemindent=0em, labelsep=0.5em]
	\item 75\% Surrogates + 25\% True Cardinalities (75\% ${Y_{s}}$ + 25\% ${Y}$): Introducing a portion of true cardinalities ${Y}$ into the model's input leads to a subtle shift in performance. While the true positive rate experiences a slight decrease to 1343 (56.4\%), the true negative rate shows a minor improvement to 727 (30.5\%). This observation suggests that the initial incorporation of true cardinalities may introduce some complexities in the classification process, potentially due to the model needing to reconcile discrepancies between the surrogates ${Y_{s}}$ and the ground truth ${Y}$. However, the decrease in false positives to 152 (6.4\%) indicates an enhanced ability to identify sub-optimal plans, even with limited true cardinality information. The overall accuracy of 86.9\% in this scenario, although slightly lower than the surrogates-only setting, highlights the potential benefits of incorporating even a smaller fraction of true cardinalities into the model's decision-making process. This sets the stage for further exploration into how the model's performance evolves as the proportion of true cardinalities increases.

	\item 50\% Surrogates + 50\% True Cardinalities (50\% ${Y_{s}}$ + 50\% ${Y}$): With a balanced mix of surrogates ${Y_{s}}$ and true cardinalities ${Y}$, the model's performance further improves, showcasing the benefits of incorporating more accurate cardinality information. The number of true positives increases to 1346 (56.6\%), and true negatives rise to 734 (30.8\%), indicating enhanced identification of both optimal and suboptimal plans. Simultaneously, we observe a reduction in both false positives to 145 (6.1\%) and false negatives to 155 (6.5\%). This translates to an overall classification accuracy of 87.39\%. This configuration demonstrates a clear advantage over the surrogates-only setting, highlighting the positive impact of integrating true cardinalities on the model's ability to correctly classify query plans, particularly in discerning suboptimal ones.

	\item 25\% Surrogates + 75\% True Cardinalities (25\% ${Y_{s}}$ + 75\% ${Y}$): Further increasing the proportion of true cardinalities in the model's input leads to a notable improvement in performance. The true positive rate rises to 1379 (57.9\%), and the true negative rate also increases to 749 (31.5\%), indicating enhanced identification of both optimal and suboptimal plans. Simultaneously, we observe a reduction in both false positives to 130 (5.5\%) and false negatives to 122 (5.1\%). This translates to an overall classification accuracy of 89.4\%. This configuration showcases the model's ability to leverage the increased accuracy of the cardinality information, leading to even better discrimination between optimal and suboptimal plans. The reduction in both types of misclassifications underscores the positive impact of incorporating a higher proportion of true cardinalities into the model's decision-making process.
\end{itemize}

\noindent
\textbf{True Cardinalities Only.}
Finally, when the model operates exclusively on true cardinalities $Y$, it achieves optimal performance, underscoring the critical role of accurate cardinality information in precise classification. The number of true positives further increases to 1410 (59.2\%), and true negatives rise to 780 (32.8\%). Concurrently, we observe a reduction in both false negatives to 91 (3.8\%) and false positives to 99 (4.2\%), leading to the highest overall classification accuracy of 91.93\% across all scenarios. This configuration exemplifies the model's ability to attain peak performance when provided with ground-truth cardinality information. As PLANSIEVE progressively incorporates true cardinalities Y, gleaned from query executions, into its cache, the model's reliance on less accurate surrogate estimates ${Y_{s}}$ diminishes. This enables the model to leverage increasingly precise cardinality information, leading to enhanced discrimination between optimal and suboptimal plans and, consequently, a notable improvement in classification accuracy.

\subsection{Effect of Cardinality Initialization} 
To assess the robustness of PLANSIEVE and its sensitivity to the accuracy of the initial surrogate cardinalities, we extend our online evaluation to include alternative third-party estimators. In addition to DeepDB, we consider two scenarios:

\begin{itemize}[leftmargin=1em, labelindent=1em, itemindent=0em, labelsep=0.5em]
	\item \textbf{Random Estimator (Rand-Est): } This estimator randomly assigns cardinalities to subqueries, simulating a scenario with highly inaccurate initial estimations.

	\item \textbf{Reversed True Cardinality (Reversed-TC): } Subqueries are sorted based on their true cardinalities, and then the cardinality values are reversed. This simulates a worst-case scenario where the relative order of subqueries based on estimated cardinalities is completely inverted compared to the true cardinality order. For instance, if  $mk \Join t$ has a cardinality of 1 and $mk \Join ci$ has a cardinality of 10, after sorting and reversing, the assigned cardinalities would be 10 for $mk \Join t$ and 1 for $mk \Join ci$.
\end{itemize}

The motivation behind these scenarios is to evaluate PLANSIEVE's performance under conditions of extreme estimation error and assess its ability to recover and refine its predictions as it incorporates true cardinalities observed during query processing. This analysis provides valuable insights into the framework's resilience and adaptability in challenging environments with unreliable initial cardinality estimates.

\begin{table*}[htbp]
    \centering
    \begin{tabular}{llrr}
    \toprule
    \textbf{Dataset} & \textbf{Estimator} & \textbf{Overall Accuracy} & \textbf{Sub-optimal Plan Accuracy} \\ \midrule
    \multirow{3}{*}{JOB-LIGHT-SCALE} & DeepDB & 93.1\% - 95.5\% & 79.3\% - 85.5\% \\ 
     & Rand-Est & 83.4\% - 95.5\% & 60\% - 85.5\%\\ 
     & Reversed-TC & 81.7\% - 95.5\% & 70\% - 85.5\% \\ 
	\midrule
	\multirow{3}{*}{STATS-CEB-SCALE} & DeepDB & 87.2\% - 92\% & 82.3\% - 88.7\% \\ 
     & Rand-Est & 77\% - 92\% & 66.8\% - 88.7\%\\ 
     & Reversed-TC & 75.8\% - 92\% & 57.7\% - 88.7\% \\ 
    \bottomrule 
    \end{tabular}
	\caption{PLANSIEVE performance with different estimators across JOB-LIGHT-SCALE and STATS-CEB-SCALE.}
    \label{tab:diff-estimators}
\end{table*}

\textbf{Impact of Initial Cardinality Estimates.}
Table \ref{tab:diff-estimators} summarizes PLANSIEVE's overall accuracy and sub-optimal plan prediction performance across two datasets—JOB-LIGHT-SCALE and STATS-CEB-SCALE—using three third-party estimators: DeepDB, Rand-Est, and Reversed-TC. The overall accuracy is presented as a range, with the lower bound reflecting performance with surrogate cardinalities and the upper bound showing improvement when true cardinalities are available.

For the \textbf{JOB-LIGHT-SCALE} dataset, DeepDB achieves an overall accuracy of 93.1\% to 95.5\%, with sub-optimal plan prediction accuracy improving from 79.3\% to 85.5\%. This indicates that DeepDB's initial estimates are good enough to maintain the relative order of sub-plans close to the optimal order. In contrast, Rand-Est and Reversed-TC, which simulate highly inaccurate estimations, exhibit lower initial accuracies of 83.4\% and 81.7\%, respectively. However, both reach up to 95.5\% with true cardinalities. The sub-optimal plan prediction accuracy for Rand-Est improves significantly from 60\% to 85.5\%, while Reversed-TC shows a substantial recovery, improving from 70.8\% to 85.5\%. These results highlight PLANSIEVE's ability to recover from poor initial estimates and refine them as more queries are processed.

In the \textbf{STATS-CEB-SCALE} dataset, DeepDB maintains robust performance, achieving an overall accuracy of 87.2\% to 92\%, with sub-optimal plan prediction accuracy improving from 82.3\% to 88.7\%. Rand-Est and Reversed-TC begin with lower overall accuracies (77\% and 75.8\%), but both eventually reach 92\% when true cardinalities are available. Sub-optimal plan prediction accuracy for Rand-Est improves from 66.8\% to 88.7\%, while Reversed-TC improves from 57.7\% to 88.7\%.

These results highlight the advantages of starting with a more accurate estimator like DeepDB, which offers superior initial performance compared to Rand-Est or Reversed-TC. However, PLANSIEVE consistently demonstrates its capacity to improve prediction accuracy, even when starting with poor estimators, by progressively refining cardinality estimates during query processing.

\subsection{Generalization to Unseen Workloads}

To evaluate PLANSIEVE's generalization ability, we conducted experiments using the JOB-LIGHT-RANGES workload \cite{Yang:NeuroCard:pvldb-2021}, which extends JOB-LIGHT with more complex range predicates. This presents a more challenging scenario, allowing us to assess PLANSIEVE's performance on unseen queries that differ from its training distribution.

\begin{table*}[htbp]
    \centering
    \begin{tabular}{lrr}
    \toprule
    \textbf{Estimator} & \textbf{Overall Accuracy} & \textbf{Sub-optimal Plan Accuracy} \\ \midrule
    NeuroCard & 72.6\% - 76.2\% & 48\% - 52.5\% \\ 
    \bottomrule 
    \end{tabular}
	\caption{PLANSIEVE performance on JOB-LIGHT-ranges when trained on JOB-LIGHT-SCALE.}
    \label{tab:job-light-range}
\end{table*}

As shown in Table \ref{tab:job-light-range}, PLANSIEVE achieves an overall accuracy of 72.6\% to 76.2\% when using NeuroCard as the cardinality estimator. However, its sub-optimal plan prediction accuracy is reduced to 48\% from 52.5\%. This degradation suggests that PLANSIEVE's ability to detect sub-optimal plans is impacted when encountering workloads significantly different from the training data.

This performance reduction can be attributed to the increased complexity of range predicates in JOB-LIGHT-RANGES, which introduces greater variability in cardinality estimates and alters the relative order of subplans. This highlights the potential need for retraining or fine-tuning to maintain high accuracy when query workloads evolve and deviate substantially from the original training set.

\subsection{Summary}
The evaluation of PLANSIEVE demonstrates its adaptability and robustness under varying conditions of cardinality availability. In the offline phase, PLANSIEVE's classification model (CM), trained with true cardinalities, outperforms the baseline in identifying sub-optimal plans, effectively minimizing false positives.  Online, PLANSIEVE begins with surrogate cardinalities, refining them as true cardinalities are observed during query execution. Even with less accurate initial cardinalities, PLANSIEVE significantly improves sub-optimal plan prediction accuracy, highlighting its resilience and ability to refine estimates.
Evaluation on the JOB-LIGHT-RANGES workload shows reduced accuracy in sub-optimal plan identification, indicating the potential need for retraining or fine-tuning when encountering significantly different workloads. Overall, PLANSIEVE proves to be a robust framework for identifying sub-optimal plans in real-time, capable of adapting to dynamic environments and refining inaccurate estimates.  Future enhancements may further improve its generalization to complex or unfamiliar query patterns.

\section{RELATED WORK}\label{sec:related-work}   

While cardinality estimation errors are unavoidable, research has focused on identifying and mitigating their impact. One significant consequence of such errors is the selection of sub-optimal query execution plans. The standard metric for quantifying individual cardinality estimation errors is Q-error \cite{Moerkotte:bad-plans}. However, its connection to plan sub-optimality is primarily through a loose worst-case upper bound of P-error \cite{Han:CEB:2021, Negi:FlowLoss:pvldb-2021, Li:Q-error-bound}, as demonstrated by Izenov et al. \cite{Izenov:L1-ERROR:sigmod-2024}. They introduced the L1-error metric, which calculates the aggregated difference between the relative order of subplans based on estimated and true cardinalities. In contrast to prior work, PLANSIEVE directly utilizes the positional vectors of subplans as input, emphasizing the importance of learning the discrepancies between estimated and true cardinality-based orderings.

Recent work by Xiu et al.\cite{Xiu:PARQO:arxiv-2024} has explored learning the distribution of Q-error for relevant subplans from historical execution data and using that error distribution to flag and avoid plans likely to produce high errors. However, we argue that choosing a set of subplans based solely on their error probability does not necessarily prevent sub-optimal plan selection. Izenov et al. \cite{Izenov:L1-ERROR:sigmod-2024} showed that even with fluctuating Q-errors in subplans, the optimizer can still find the optimal plan if the relative order of subplans remains consistent with the optimal one. PLANSIEVE adheres to this principle by focusing on the relative ordering of subplans rather than individual error probabilities.

The existing literature also explores robust query optimization (RQO) techniques to address the inherent uncertainty in cardinality estimations \cite{Borovica:RQO:Dagstuhl, Goetz:RQO:Dagstuhl, Haritsa:RQO:ICDE}. Approaches such as LEC \cite{Francis:LEC:SIGMOD} and RCE \cite{Babcock:RCE:SIGMOD} employ probability distributions and sampling to estimate selectivity and identify plans with the lowest expected cost. Interval-based methods like Rio \cite{Babu:Rio:SIGMOD} and its extensions \cite{Belgin:Rio-ext:PDCAT} utilize bounding boxes or intervals to quantify selectivity uncertainty, while adaptive processing techniques  \cite{Dutt:plan-bouquet:SIGMOD, Dutt:QUEST:PVLDB, Trummer:SKINNERDB:sigmod-2019} adjust query plans during execution based on runtime observations. However, these methods often rely on strong assumptions about selectivity distributions, require runtime adaptations, or necessitate extensive offline training. Plan diagrams \cite{Harish:PD:VLDB-2007, Harish:PD:VLDB-2008} and their subsequent refinements \cite{Dey:PD:PVLDB, Purandare:RQO:TR-2018} aim to identify robust plans for query templates across the selectivity space.

In contrast to these approaches, PLANSIEVE focuses on the online identification of sub-optimal plans during query optimization. While PLANSIEVE operates online, it still requires an initial offline training phase to build its classification model. The core of the PLANSIEVE classification model is a transformer architecture. Previous applications of transformers in this domain have primarily focused on predicting the cardinality of selections or joins between tables \cite{Yang:naru:VLDB-2019,Yang:NeuroCard:pvldb-2021, Tsan:sketch:sigmod24}, or predicting joinability across tables \cite{Nobari:DTT}. In contrast, PLANSIEVE's classification model employs a transformer to learn the relative differences in subplan orderings.

\section{CONCLUSIONS}\label{sec:conclusions}

This work introduces PLANSIEVE, a novel proactive approach to address the critical challenge of cardinality estimation errors during query optimization. By leveraging surrogate true cardinalities and relative subplan order analysis, PLANSIEVE provides a practical solution for online identification of sub-optimal plans. Our empirical evaluation demonstrates PLANSIEVE's effectiveness in enhancing the accuracy of sub-optimal plan detection as the system processes more queries. Integration of PLANSIEVE into conventional query optimizers offers the potential to reduce the negative impact of cardinality estimation inaccuracies.
Future research can explore potential actions to be taken when PLANSIEVE flags a sub-optimal plan, as well as investigating how to utilize true cardinality-based position vectors to identify robust execution plans.

\paragraph*{Acknowledgments.} This work is supported by NSF award number 2008815.

\bibliographystyle{abbrv}

\end{document}